\documentclass{article}

\PassOptionsToPackage{numbers, compress}{natbib}

\usepackage[main,final]{neurips_2026}

\usepackage{amsmath}
\usepackage{amssymb}
\usepackage{mathtools}
\usepackage{amsthm}
\usepackage{multirow}
\usepackage{microtype}
\usepackage{graphicx}
\usepackage{subcaption}
\usepackage{booktabs} 
\usepackage{hyperref}
\usepackage{algorithm}
\usepackage{algpseudocode}
\usepackage{xcolor}

\theoremstyle{plain}
\newtheorem{theorem}{Theorem}[section]
\newtheorem{problem}{Problem}

\newtheorem{lemma}[theorem]{Lemma}
\newtheorem{corollary}[theorem]{Corollary}
\theoremstyle{definition}
\newtheorem{definition}[theorem]{Definition}

\theoremstyle{remark}

\newcommand{\N}{\mathbb{N}}

\newcommand{\R}{\mathbb{R}}
\newcommand{\Sys}{\mathfrak{S}}
\newcommand{\Xx}{\mathcal{X}}

\newcommand{\Relu}{\mathsf{ReLU}}
\newcommand{\vect}[1]{\mathbf{#1}}

\newcommand{\real}{\mathbb{R}}

\newcommand{\setdef}[2]{\{#1
	\; | \; #2\}}

\usepackage{enumitem}
\usepackage[utf8]{inputenc} 
\usepackage[T1]{fontenc}    
\usepackage{hyperref}       
\usepackage{url}            
\usepackage{booktabs}       
\usepackage{amsfonts}       
\usepackage{nicefrac}       
\usepackage{microtype}      
\usepackage{xcolor}         

\title{Scalable Neural Safety Certification via Monotonicity}

\author{
  Amirreza Alavi \\
  University of Colorado Boulder \\
  Boulder, CO, USA \\
  \texttt{seyedamirreza.alavi@colorado.edu}
  \And
  Majid Zamani \\
  University of Colorado Boulder \\
  Boulder, CO, USA \\
  \texttt{majid.zamani@colorado.edu}
  \And
  Saber Jafarpour \\
  University of Colorado Boulder \\
  Boulder, CO, USA \\
  \texttt{saber.jafarpour@colorado.edu}
}

\begin{document}

\maketitle
\begin{abstract}
Learning-based safety certification offers a promising route for verifying
complex dynamical systems when explicit models are unavailable, but existing
data-driven methods often scale poorly: they treat the system as a black box,
rely on dense discretizations or Lipschitz bounds, and consequently suffer from
sample complexity that grows exponentially with dimension. This paper shows that
a common structural property of dynamical systems---\emph{monotonicity}---can be used
to break this curse of dimensionality. We develop a data-driven framework for
robust safety verification and safe controller synthesis for unknown monotone
systems using only simulator queries. Our main theoretical result proves that,
for monotone systems with upper-closed unsafe sets, safety is completely
characterized by the existence of monotone inductive barrier certificates. This
structure reduces certificate verification over continuous state spaces to
localized boundary checks over finitely many cells. To learn such certificates,
we introduce Max-Monotone Networks, a monotone neural architecture with
universal approximation guarantees for monotone functions. We then propose a
training and refinement algorithm that, upon successful termination, returns
formally valid neural barrier certificates and, under mild partition assumptions,
requires only $O(n)$ simulator queries. Across high-dimensional benchmarks,
including oscillator networks with up to \(13{,}659\) states and traffic
networks with \(1{,}000\) states, our method certifies safety in regimes where
prior approaches either do not converge or exceed memory limits. These results
demonstrate that exploiting order structure can make neural safety certification
both scalable and formally sound.

\end{abstract}

\section{Introduction}

Dynamical systems provide a powerful mathematical framework for modeling and
analyzing autonomous, engineered, and natural systems. They are widely used to
describe complex behaviors arising in safety-critical domains such as
self-driving vehicles, biological networks, and industrial process control. As
autonomous systems become increasingly prevalent, it is essential to develop
rigorous methods that can certify their safe operation over long or infinite time
horizons. Barrier certificates provide a powerful inductive approach for this
purpose: they reduce safety verification to the search for a scalar function
that separates initial states from unsafe states and is preserved along system
trajectories~\cite{prajna2004safety,ADA-SC-ME-GN-KS-PT:19}. When such a
function exists, it serves as a formal certificate that the unsafe set is
unreachable.

Although barrier certificates provide a clean inductive proof of safety, they
remain difficult to compute for high-dimensional systems, especially in
data-driven settings where the dynamics are unknown and only simulator-generated
data or trajectories are available. Classical barrier-certificate methods
construct certificates using linear programming~\cite{SS-XC-EA:13,YZ-etal:16},
sum-of-squares (SOS) optimization~\cite{prajna2004safety}, or
satisfiability-modulo-theories (SMT) techniques~\cite{edwards2024fossil}. These
approaches have led to substantial progress and provide strong formal
guarantees~\cite{prajna2006barrier,ADA-XX-JWG-PT:17,PW-FA:07}, but they often
require explicit models, restrictive dynamics, or optimization procedures that
scale poorly with dimension. Neural networks have recently emerged as a flexible
representation class for learning barrier certificates from
data~\cite{chang2019neural,zhao2020synthesizing,zhang2024exact,robey2020learning}.
However, existing neural barrier frameworks typically rely on Lipschitz bounds,
dense sampling, or expensive verification subroutines to obtain formal
guarantees. As a result, their sample complexity and computational cost can grow
rapidly with the state dimension.

This paper shows that a fundamental structural property of dynamical
systems---\emph{monotonicity}---can be used to overcome this scalability bottleneck.
Monotonicity means that the system preserves a natural partial order on states
and inputs throughout its evolution. This property has been widely studied in dynamical systems and formal methods as
a way to capture cooperative behavior and monotone structure in complex
systems~\cite{FINKEL200163,Blondin2016WellBT}. Monotone dynamics arise
naturally in many real-world settings, including traffic networks~\cite{CFD:94,SC-MA:15}, queuing systems~\cite{LF-SR:11}, population
dynamics~\cite{leenheer2004predator}, chemical reaction networks~\cite{DA:13},
and biological systems~\cite{EDS:07}. In many of these settings, monotonicity
follows \emph{naturally} from the physical or network structure of the system. Moreover,
monotonicity has also emerged as a \emph{design principle} for ensuring closed-loop
performance and robustness~\cite{liu2020certified,DA-EDS:03}. Even when a
system is not inherently monotone, controllers can often be designed to enforce
monotonicity in the closed-loop dynamics, as in droop control for power
systems~\cite{JWS-FD-FB:13}, consensus protocols in multi-agent
robotics~\cite{HB-MA-JW:11}, and structural control strategies in biological
networks~\cite{DA-EDS:03,EDS:07}.

A central advantage of monotone dynamics is that order preservation makes many
global questions reducible to extremal cases over order intervals, a principle
that underlies scalable analysis and control methods for this class of
systems~\cite{HLS:95,DA-EDS:03}. These properties have been exploited to
establish stability of interconnected systems~\cite{DA-EDS:03}, enable
efficient control synthesis~\cite{AR:15,YE-DP-DA:17}, and support scalable
reachability analysis~\cite{SC-MA:15b}. Despite this rich body of work, the role
of monotonicity in data-driven safety certification remains much less developed,
with only limited recent progress on neural barrier certificates for monotone
systems~\cite{AA-etal:25}. In particular, it has remained unclear whether
monotonicity can provide a complete barrier-certificate characterization of
safety, whether such certificates can be learned from simulator queries alone,
and whether the resulting verification procedure can avoid the exponential
sample complexity typical of black-box methods.

Our key observation is that monotonicity turns the certificate-learning problem
from a global search over the state space into a finite set of boundary
conditions. For monotone systems with upper-closed unsafe sets, we prove a
complete characterization: safety holds if and only if there exists a \emph{monotone}
inductive barrier certificate. Moreover, the monotonicity of the certificate and
dynamics allows global safety constraints over continuous state and input spaces
to be reduced to localized boundary checks over finitely many cells. Thus,
instead of densely sampling the full state space, safety certification can be
performed using carefully selected simulator queries.

Building on this result, we develop a learning-based framework for robust safety
verification and safe controller synthesis for unknown monotone systems. We
introduce \emph{Max-Monotone Networks}, a monotone neural architecture with
universal approximation guarantees for monotone functions. The architecture is
designed so that the learned neural certificate respects the order structure
required by our theory. We then propose a training and refinement algorithm that
uses simulator queries to learn neural barrier certificates satisfying the
finite boundary conditions. Upon successful termination, the algorithm returns a
formally valid certificate of safety. Under mild partition assumptions, the
number of simulator queries required by the refinement procedure scales
\emph{linearly} in the system dimension.

Our contributions are summarized as follows:
\begin{itemize}
    \item We prove that, for monotone discrete-time systems with upper-closed
    unsafe sets, robust safety verification and safe controller synthesis admit
    complete characterizations via \emph{monotone} inductive barrier certificates.

    \item We show that monotonicity reduces global certificate verification over
    continuous state and input spaces to finite localized boundary checks,
    enabling formal safety certification from simulator queries.

    \item We introduce \emph{Max-Monotone Networks}, a monotone neural architecture
    with universal approximation guarantees, tailored to learning monotone
    barrier certificates.

    \item We develop a training and adaptive refinement algorithm that returns
    formally valid neural barrier certificates upon successful termination and
    achieves \(O(n)\) simulator-query complexity under mild partition
    assumptions.

    \item We demonstrate scalability on high-dimensional benchmarks, including
    traffic-flow models, Kuramoto oscillator networks, power-system models,
    Lotka--Volterra dynamics, and ribosome-flow models. Our method certifies
    systems with thousands of dimensions, including a \(13{,}659\)-state
    oscillator network and a \(1{,}000\)-state traffic network, while competing
    approaches either fail to converge or exceed memory limits.
\end{itemize}

Together, these results show that exploiting order structure can turn neural
safety certification from a high-dimensional black-box sampling problem into a scalable, formally certified learning procedure.

\noindent\textbf{Code availability.}
The implementation and experimental code are available in our
\href{https://github.com/seyedamirrezaalavi/Scalable-Neural-Safety-Certification-via-Monotonicity}{GitHub repository}.

\section{Preliminaries and problem statement}

In this section, we review the necessary preliminaries and formulate the main problem studied in this paper. 

\paragraph{Mathematical Notations.}
We denote the sets of real, nonnegative real, and positive real numbers by $\R$, $\R_{\ge 0}$, and $\R_{>0}$, respectively, and the sets of nonnegative integers and positive integers by $\N$ and $\N_{>0}$, respectively. Given an integer $K\in \mathbb{N}_{>0}$, we define $[K]=\{0,1,\ldots,K\}$. Let $\mathcal{X} \subseteq \R^n$ and $f:\mathcal{X} \to \mathcal{X}$. For $i\in \mathbb{N}$, we define $f^{i}:\mathcal{X}\to \mathcal{X}$ recursively by $f^{i}(x) = f(f^{(i-1)}(x))$, for every $i\in \mathbb{N}_{>0}$ and $f^{0}(x) = x$, for every $x\in \mathcal{X}$. Let $\mathcal{X} \subseteq \R^n$, $U \subseteq \R^m$, and $f:\mathcal{X}\times U \to \mathcal{X}$. For an input sequence $\mathbf{u} = (u_0,\ldots,u_{K-1})\in U^{K}$, we define the iterates of $f$ recursively as
\begin{align*}
    f^{0}(x,\mathbf{u}) &= x, && \forall \; x\in \Xx\\
    f^{i+1}(x,\mathbf{u}) &= f\big(f^{i}(x,\mathbf{u}),u_{i}\big),
    \quad &&\forall \; x\in \Xx,\;  \forall \; i\in [K-1].
\end{align*}
For a feedback controller $\kappa:\Xx\to U$, we use $f^k(x,\kappa(x))$ to denote the
$k$-step closed-loop evolution under
$x_{t+1}=f(x_t,\kappa(x_t))$, with $x_0=x$. For sets $A$ and $B \subseteq A$, we use $A \setminus B$ for set difference, $A \times B$ for Cartesian product, and write $|A|$ for the cardinality of $A$. For $k \in \mathbb{N}_{\ge 0}$, we denote by $A^{k}$ the $k$-fold Cartesian product of the set $A$ with itself.
For a vector $x = (x_1,\ldots,x_n)^\top \in \R^n$, we denote its $i$th component by $x_i$. We write $x \le y$ to indicate element-wise inequality, i.e., $x_i \le y_i$ for all $i \in \{1,\ldots,n\}$. The Euclidean norm and the $\ell_1$-norm of $x$ are denoted by $\|x\| = \sqrt{\sum_{i=1}^n x_i^2}$ and $\|x\|_1 = \sum_{i=1}^n |x_i|$, respectively.
A hyper-rectangle is represented by $[\underline{x}, \overline{x}] = \{ x \in \R^n \mid \underline{x} \le x \le \overline{x} \}$. A set $S\subseteq\mathcal{X}$ is \emph{lower closed} (resp. \emph{upper closed}) relative to $\mathcal{X}$ if $x\in S$ and $y\in\mathcal{X}$ with $y\le x$ (resp. $y\ge x$) imply $y\in S$. Given a set $A \subseteq \R^n$, a point $x \in A$ is \emph{maximal} (resp. \emph{minimal}) if there is no $y \in A$ such that $y\ne x$ and $x \le y$ (resp. $y \le x$).
Let $\mathcal{X} \subseteq \R^n$. A map $f:\mathcal{X} \to \mathcal{Y}$ is \emph{monotone} if, for all $x, y \in \mathcal{X}$ such that $x \le y$, we have $f(x) \le f(y)$.
The rectified linear unit (ReLU) is defined by $\mathrm{ReLU}(x) = \max(0, x)$. 
Throughout the paper, $0_{n \times n}$ denotes the $n \times n$ zero matrix, and $\vect{1}_n$ denotes the $n$-dimensional column vector of ones. 

\subsection{Dynamical systems}

Dynamical systems provide a fundamental framework for describing the evolution of natural or engineered processes over time. 
\begin{definition}[Dynamical systems]
\label{def: dts}
A \emph{discrete-time dynamical system (dtDS)} is a tuple $\mathfrak{S} \coloneqq (\mathcal{X},\mathcal{X}_0,U,f)$, where $\mathcal{X}\subseteq \R^n$ denotes the set of states, $\mathcal{X}_0 \subseteq \mathcal{X}$ is the set of initial states, $U\subseteq \R^m$ is the set of inputs, and $f:\mathcal{X} \times U \to \mathcal{X}$ is the state-transition map.
\end{definition}
For an input sequence $\{u_t\}_{t=0}^{\infty} \in U$, the evolution of the system $\Sys$ is given by
\begin{align}\label{eq:mon}
  x_{t+1} = f(x_t,u_t),\qquad \forall t\ge 0.
\end{align}
The resulting state sequence $\{x_t\}_{t=0}^{\infty}$ is called the \textit{trajectory} of $\Sys$ associated with the input $\{u_t\}_{t=0}^{\infty}$. 

Safety is a fundamental property of dynamical systems. Given a dynamical system $\Sys$, we adopt a set-theoretic perspective on its safety by specifying a set $\mathcal{X}_u \subseteq \mathcal{X}$ of unsafe states. The system $\Sys$ is safe if its state remains outside $\mathcal{X}_u$ for all time.

\begin{definition}[Safety of dynamical systems]\label{def:safety}
Consider a dtDS $\Sys = (\Xx, \Xx_0, U, f)$ and a set of unsafe states $\Xx_u \subseteq \Xx$. Then
\begin{enumerate}[label=\arabic*)]

\item (\textit{Safe controllability}) the system $\Sys$ is \textit{safely controllable} with respect to $\Xx_u$ if, for every initial state $x_0 \in \Xx_0$, there exists a control sequence $\{u_t\}_{t=0}^{\infty} \in U$ such that the corresponding trajectory $\{x_t\}_{t=0}^{\infty}$ starting from $x_0$ satisfies $\{x_t\}_{t=0}^{\infty}\cap \Xx_u=\emptyset$. 

\item (\emph{Robust safety}) the system $\Sys$ is \textit{robustly safe} with respect to $\Xx_u$ if, for every initial state $x_0 \in \Xx_0$ and every input sequence $\{u_t\}_{t=0}^{\infty} \in U $, the corresponding trajectory $\{x_t\}_{t=0}^{\infty}$ starting from $x_0$ satisfies $\{x_t\}_{t=0}^{\infty}\cap \Xx_u=\emptyset$.

\end{enumerate}
\end{definition}
The distinction between safe controllability and robust safety lies in the
interpretation of the input \(U\) in the dtDS \(\Sys\).
For safe controllability, inputs represent control actions, and the goal is to
\emph{design} a policy \(\{u_t\}_{t=0}^{\infty} \in U\) that keeps the
closed-loop system safe.
For robust safety, inputs represent disturbances, and safety must hold for
\emph{all} admissible input sequences \(\{u_t\}_{t=0}^{\infty} \in U\).

\subsection{Safety via inductive barrier certificates}
\label{sec:barrier}

A broad class of inductive methods, known as \emph{barrier certificates}, has
been developed for safety analysis of dynamical
systems~\cite{prajna2004safety,ADA-XX-JWG-PT:17,SB:2018,AnandKinductive2021}.
Intuitively, a barrier certificate separates the initial set from the unsafe set
and inductively proves that all system trajectories avoid unsafe states.
Consider a dtDS \(\Sys = (\Xx, \Xx_0, U, f)\) with unsafe set
\(\Xx_u \subseteq \Xx\).
A function $\mathcal{B}:\mathcal{X}\to \mathbb{R}$ is a \emph{control $K$-inductive barrier certificate} for $\Sys$ with respect to $\Xx_u$ if there exist $\gamma,\delta,\eta \in \mathbb{R}$ such that $\gamma \le \delta < \eta$, and
\begin{align}
\max_{k\in [K-1]}\; &\mathcal{B}\!\left(f^{k}(x,\mathbf{u})\right) \le \gamma && \forall \; x \in \Xx_0 \;, \exists\; \mathbf{u}\in U^{K-1},\label{eq:indbar1-c}\\
&\mathcal{B}(x) \geq \eta &&\forall \; x \in \Xx_u,\label{eq:indbar2-c}
\end{align}
and, for every $x\in\Xx$ and every $\mathbf{u}\in U^{K-1}$, there exists $u\in U$ such that
\begin{align}\label{eq:indbar3-c}
\max_{k\in [K-1]} \mathcal{B}\!\left(f^{k}(x,\mathbf{u})\right)\le \delta 
\implies \mathcal{B}\!\left(f(f^{K-1}(x,\mathbf{u}),u)\right) \le \delta.
\end{align}
We refer to mathematical notation section for the definitions of $f^{k}(x,\mathbf{u})$ and $[K-1]$. Similarly, a function $\mathcal{B}:\mathcal{X}\to \mathbb{R}$ is a \emph{robust $K$-inductive barrier certificate} for $\Sys$ with respect to $\Xx_u$ if there exist $\gamma,\delta,\eta \in \mathbb{R}_{\ge 0}$ such that $\gamma \le \delta < \eta$ and, for every $\mathbf{u}\in U^{K}$, the following conditions hold:
\begin{align}
\max_{k\in [K-1]}\; &\mathcal{B}\!\left(f^{k}(x,\mathbf{u})\right) \le \gamma && \forall \; x \in \Xx_0,\label{eq:indbar1}\\
&\mathcal{B}(x) \ge \eta &&\forall \; x \in \Xx_u,\label{eq:indbar2}\\
\max_{k\in [K-1]} &\mathcal{B}\!\left(f^{k}(x,\mathbf{u})\right)\le \delta
\implies \mathcal{B}\!\left(f^{K}(x,\mathbf{u})\right) \le \delta, &&\forall \; x \in \Xx.
\label{eq:indbar3}
\end{align}

When \(K=1\), control and robust \(1\)-inductive barrier certificates reduce to
the classical control and robust barrier certificates studied in the control
literature~\cite{ADA-XX-JWG-PT:17,prajna2004safety}.
The following standard result states that the existence of such a certificate
guarantees safety. We refer to~\cite{prajna2004safety,ADA-XX-JWG-PT:17,VM:2024} for a proof.

\begin{theorem}[Inductive barrier safety certificates]\label{thm:barrier}
    Consider a dtDS $\Sys = (\Xx, \Xx_0, U, f)$ with a set of unsafe states $\Xx_u \subseteq \Xx$ and an integer $K\in \mathbb{N}_{\ge 1}$. Then $\mathfrak{S}$ is 
    \begin{enumerate}[label=\arabic*)]
         \item\label{p2:barrier} safely controllable if and only if there exists a control $K$-inductive barrier function $\mathcal{B}:\mathcal{X}\to \mathbb{R}$ for $\Sys$ with respect to $\Xx_u$. 
        \item\label{p1:barrier} robustly safe if and only if there exists a robust $K$-inductive barrier function $\mathcal{B}:\mathcal{X}\to \mathbb{R}$ for $\Sys$ with respect to $\Xx_u$. 
        \end{enumerate}
\end{theorem}

Theorem~\ref{thm:barrier} reduces robust safety, or safe
controllability, to finding a function
\(\mathcal{B}:\mathcal{X}\to\mathbb{R}\) satisfying
conditions~\eqref{eq:indbar1-c}--\eqref{eq:indbar3-c}, or
conditions~\eqref{eq:indbar1}--\eqref{eq:indbar3}.
This functional characterization enables the use of computational tools for
searching for \(K\)-inductive barrier certificates, including SMT
solvers~\cite{edwards2024fossil} and SOS
methods~\cite{prajna2004safety}.

\subsection{Monotone dynamical systems}
We define the class of monotone systems characterized by order-preserving state-transition maps.

\begin{definition}[Monotone Dynamical Systems]\label{def:monsys}
A dtDS $\Sys = (\mathcal{X}, \mathcal{X}_0, U, f)$ is \emph{monotone} if the map $f : \mathcal{X} \times U \to \mathcal{X}$ satisfies
\[
 x \le y \;\;\mbox{ and }\;\; u\le v  \quad\implies\quad  f(x,u) \le f(y,v).
\]
\end{definition}
It can be shown that, when the  map $f$ is differentiable and the sets $\mathcal{X}$ and $U$ are convex, the discrete-time system $\Sys$ is monotone if and only if $Df(x,u) \ge 0_{n \times (n+m)}$ for all $(x,u) \in \mathcal{X} \times U$, where $Df(x,u)$ denotes the Jacobian of $f$ with respect to $(x,u)$~\cite{HLS:95}.

As discussed in the introduction, monotone systems arise in many natural and
engineering applications, either because the underlying physics is
\emph{inherently} monotone or because monotonicity is deliberately enforced in
\emph{system design} to enable tractable analysis and control. Nevertheless,
many important real-world systems are non-monotone due to competitive,
inhibitory, or antagonistic interactions among state variables.

\textit{Monotone Embeddings.} Studying monotone systems entails limited loss of generality. A fundamental
observation in monotone dynamical systems is that any dynamical system can be
\emph{embedded} into a higher-dimensional monotone
system~\cite{GAE-HLS-EDS:06,HLS:08,SC:20,SC-MA:15b} by separating cooperative
and competitive interactions among state variables.
Thus, although monotonicity is a structural assumption, it is not a narrow one:
monotone systems can represent general nonlinear dynamics through suitable
embeddings.
Consequently, analytical and computational tools for monotone systems can often
be extended to broad classes of non-monotone systems via such embeddings.

\subsection{Problem statements}

In this paper, we study monotone systems with \emph{unknown} state-transition
maps. We assume access to a simulator queried at arbitrary state--input pairs to
evaluate the transition map. Our goal is to use these queries to establish safe
controllability and robust safety, leading to the following two problems.

\begin{problem}[Safe control synthesis]\label{p2}
Consider a monotone dtDS $\Sys = (\Xx, \Xx_0, U, f)$ with an unknown map $f$ and a set of unsafe states $\Xx_u \subseteq \Xx$. Given a finite collection of simulator data $\{f(x^i,u^i)\}_{i=1}^{N}$ corresponding to sampled states $\{x^i\}_{i=1}^{N} \in \Xx$ and sampled control actions $\{u^i\}_{i=1}^{N} \in U$, design a feedback controller $\kappa: \Xx \rightarrow U$ such that the states of the closed-loop system
\begin{equation*}
x_{t+1} = f(x_t, \kappa(x_t)), \quad t \ge 0,
\end{equation*}
never enters the unsafe set $\Xx_u$.
\end{problem}

\begin{problem}[Robust safety verification]\label{p1}
    Consider a monotone dtDS $\Sys = (\Xx, \Xx_0, U, f)$ with an unknown map $f$ and a set of unsafe states $\Xx_u \subseteq \Xx$. Given a finite collection of simulator data $\{f(x^i, u^i)\}_{i=1}^{N}$ corresponding to sampled states $\{x^i\}_{i=1}^{N} \in \Xx$ and the sampled disturbance $\{u^i\}_{i=1}^{N} \in U$, determine whether system $\Sys$ is robustly safe with respect to $\Xx_u$.
\end{problem}

\textit{Simulator access.} In both Problems~\ref{p2} and~\ref{p1}, we assume access to a simulator that evaluates $f(x^i,u^i)$ at any prescribed state–input pair $(x^i,u^i)$. For safe control synthesis, inputs represent control actions to be selected by the controller. For robust safety verification, inputs represent disturbances: although their realizations during system operation are uncontrolled, their values can be prescribed when querying the simulator.


\textit{Monotonicity as a structural property.} In Problems~\ref{p2} and~\ref{p1}, monotonicity is assumed without access to the
explicit transition map. This assumption is often justified by prior knowledge
of the system's qualitative behavior, inferred, for example, from physical laws
or geometric structure. In particular, Definition~\ref{def:monsys}
characterizes monotonicity as a \textit{qualitative} property: increasing states
or inputs cannot decrease the system response. This property arises naturally in
many engineered and natural systems, including traffic networks~\cite{SC-MA:15},
biological processes~\cite{EDS:07}, and supply chains~\cite{yu2019equilibrium}.

\section{Monotone inductive barrier certificates}

This section identifies a structural property of inductive barrier certificates
for monotone systems. We show that, when the unsafe set satisfies a mild
order-theoretic condition, inductive barrier certificates can be chosen to
inherit the \emph{monotonicity} of the dynamics. This observation simplifies
safety verification and control synthesis and enables scalable, data-driven
formulations.
We now state the result; the proof is given in Appendix~\ref{App:proofiff}.

\begin{theorem}[Monotone inductive barrier certificates]\label{thm:monbar}
    Consider a monotone dtDs $\Sys :=(\Xx,\Xx_0,U,f)$ with an upper closed set\footnote{A set $S\subseteq\mathcal{X}$ is upper closed
    relative to $\mathcal{X}$ if, for every $x\in S$ and $y\in\mathcal{X}$ with $x\le y$, we have $y\in S$.} of unsafe states $\Xx_u \subseteq \Xx$ 
    and an integer $K\in \mathbb{N}_{\ge 1}$. Then
    \begin{enumerate}[label=\arabic*)]
\item\label{part2} \textit{(Safe control synthesis)} $\Sys$ is safely controllable with respect to $\Xx_u$ if and only if there exists a monotone control $K$-inductive barrier certificate $\mathcal{B}$ satisfying conditions~\eqref{eq:indbar1-c},~\eqref{eq:indbar2-c}, and~\eqref{eq:indbar3-c}. 
    
\item\label{part1} \textit{(Robust safety verification)} $\Sys$ is robustly safe with respect to $\Xx_u$ if and only if there exists a monotone robust $K$-inductive barrier certificate $\mathcal{B}$ satisfying conditions~\eqref{eq:indbar1},~\eqref{eq:indbar2}, and~\eqref{eq:indbar3}.

    \end{enumerate}
\end{theorem}

\textit{Reduction to monotone certificate search.} For upper-closed unsafe sets, Theorem~\ref{thm:monbar} reduces safe control
synthesis, and robust safety verification of monotone systems to the
search for \emph{monotone} control and robust, inductive barrier
certificates. These are monotone functions satisfying
conditions~\eqref{eq:indbar1-c}--\eqref{eq:indbar3-c}, and
conditions~\eqref{eq:indbar1}--\eqref{eq:indbar3}.
This result is the main theoretical contribution of the paper and forms the
foundation of our data-driven approach to safety certification.

\textit{Upper closed unsafe sets.}  The assumption of upper-closed unsafe sets may seem restrictive, since not every
unsafe set has this form. However, in many safety-critical systems, unsafe
behavior occurs when state variables or aggregate resources exceed thresholds,
which naturally yields upper-closed unsafe sets. Examples include drug delivery
systems, where excessive chemical concentration can be harmful, and traffic
networks, where vehicle density above a critical threshold can cause congestion.

\subsection{Data-driven inductive barrier certificates}\label{sec:data-driven}

Although Theorem~\ref{thm:monbar} restricts the search for inductive barrier
certificates to monotone functions, verifying
conditions~\eqref{eq:indbar1-c}--\eqref{eq:indbar3-c} (resp.~\eqref{eq:indbar1}--\eqref{eq:indbar3}) still requires either
explicit knowledge of the transition map or infinitely many simulator queries
over the full state--input space.
Building on Theorem~\ref{thm:monbar}, we exploit certificate monotonicity to
construct safety certificates from finite samples while preserving rigorous
guarantees.
The key idea is to partition the state space and use monotonicity to reduce the
infinite-dimensional verification problem to finitely many boundary checks. We partition the state space $\mathcal{X}$ and the input space $U$ into
collections of hyper-rectangular cells, denoted by
$\{[\underline{x}^i,\bar{x}^i]\}_{i\in\mathcal{P}}$ and
$\{[\underline{u}^i,\bar{u}^i]\}_{i\in\mathcal{Q}}$, respectively, where
$\mathcal{P}$ and $\mathcal{Q}$ are finite index sets associated with these
partitions. For given subsets $\mathcal{X}_0 \subseteq \mathcal{X}$ and
$\mathcal{X}_u \subseteq \mathcal{X}$, we define index sets $\mathcal{I}$ and
$\mathcal{U}$ as the minimal subsets of $\mathcal{P}$ whose union covers
$\mathcal{X}_0$ and $\mathcal{X}_u$, respectively, namely
\begin{align}\label{eq:cover}
\mathcal{X}_0 \subseteq \bigcup_{i \in \mathcal{I}} [\underline{x}^i,\bar{x}^i],
\qquad
\mathcal{X}_u \subseteq \bigcup_{i \in \mathcal{U}} [\underline{x}^i,\bar{x}^i].
\end{align}
By construction, every state $x \in \mathcal{X}_0$ is component-wise bounded above by $\bar{x}^i$ for some $i \in \mathcal{I}$, while every state $x \in \mathcal{X}_u$ is component-wise bounded below by $\underline{x}^i$ for some $i \in \mathcal{U}$. 
The search for control barrier certificates that satisfy~\eqref{eq:indbar1-c}-\eqref{eq:indbar3-c} can be conservatively reformulated as finding monotone inductive barrier certificates $\mathcal{B}:\Xx \to \mathbb{R}$ and monotone feedback controller $\kappa: \Xx\to U$ satisfying 
\begin{align}
\max_{k\in [K-1]}\; &\mathcal{B}(f^{k}(\bar{x}^i,\kappa(\bar{x}^i))) \leq \gamma && \forall\; i \in \mathcal{I},\label{eq:data1-c}\\
&\mathcal{B}(\underline{x}^i) \geq \eta  && \forall\;i \in \mathcal{U},\label{eq:data2-c}\\
\max_{k\in [K-1]}\; &\mathcal{B}(f^{k}(\underline{x}^i,\kappa(\underline{x}^i))) \leq \delta \quad\implies \quad \mathcal{B}(f^{K}(\bar{x}^i,\kappa(\bar{x}^i)))\leq \delta, &&\forall\; i\in \mathcal{P} \label{eq:data3-c}  
\end{align} 
Conditions~\eqref{eq:data1-c}--\eqref{eq:data3-c} reduce certification of inductive barrier certificates over the continuous state set to finitely
many maximal and minimal points from the coverings of \(\Xx\) and \(U\).
Similarly, the search for robust barrier certificates satisfying~\eqref{eq:indbar1}-\eqref{eq:indbar3} can be conservatively reformulated as finding a monotone function $\mathcal{B}:\Xx \to \mathbb{R}$ such that,  
\begin{align}
\max_{k\in [K-1]}\; &\mathcal{B}(f^{k}(\bar{x}^i,\overline{\mathbf{u}})) \leq \gamma && \forall\; i \in \mathcal{I},\; \forall \overline{\mathbf{u}}\in \overline{\mathcal{U}}^{K}\label{eq:data1}\\
&\mathcal{B}(\underline{x}^i) \ge \eta && \forall\; i \in \mathcal{U},\label{eq:data2}\\
\max_{k\in [K-1]} &\mathcal{B}(f^{k}(\underline{x}^i,\underline{\mathbf{u}})) \leq \delta \Rightarrow \mathcal{B}(f^{K}(\bar{x}^i,\overline{\mathbf{u}}))\leq \delta,&& \forall i\in \mathcal{P},\; \forall \underline{\mathbf{u}}\in \underline{\mathcal{U}}^{K}, \; \forall \overline{\mathbf{u}}\in \overline{\mathcal{U}}^{K+1}\label{eq:data3}   
\end{align}
where $\underline{\mathcal{U}} = \bigcup_{j\in Q}\{\underline{u}^j\}$ and $\overline{\mathcal{U}} = \bigcup_{j\in Q}\{\overline{u}^j\}$.  Conditions~\eqref{eq:data1}–\eqref{eq:data3} reduce the verification of barrier certificates from the continuous state set to a finite set of carefully selected maximal and minimal points derived from the coverings of $\Xx$ and $U$.

The following result gives a data-driven formal guarantee for robust safety
verification and safe control synthesis from finite samples; see
Appendix~\ref{App:proofdata1} for the proof.

\begin{theorem}[Data-driven barrier certificates]\label{thm:datamonbar}
Consider a monotone dtDS $\mathfrak{S} = (\mathcal{X}, \mathcal{X}_0, U, f)$ with unknown map $f$ and a set of unsafe states $\mathcal{X}_u\subseteq \Xx$. Let $\{[\underline{x}^i, \bar{x}^i]\}_{i \in \mathcal{P}}$ be a family of hyper-rectangles covering the set $\mathcal{X}$, and let $\{[\underline{u}^j, \bar{u}^j]\}_{j \in \mathcal{Q}}$ be a family of hyper-rectangles covering the input space $U$. Define the index sets $\mathcal{I}$ and $\mathcal{U}$ as the minimal subsets of $\mathcal{P}$ satisfying \eqref{eq:cover}. Then, the system $\mathfrak{S}$
\begin{enumerate}[label=\arabic*)]
\item\label{thm:datamonbar1} \textit{(Safe control synthesis)} is safely controllable with respect to $\mathcal{X}_u$ if there exists monotone function $\mathcal{B}: \mathcal{X} \to \real$ and  $\kappa : \mathcal{X} \to U$ satisfying conditions~\eqref{eq:data1-c}--\eqref{eq:data3-c} with $\gamma\le \delta<\eta$.
\item\label{thm:datamonbar2} \textit{(Robust safety verification)} is robustly safe with respect to $\mathcal{X}_u$ if there exists a monotone function $\mathcal{B}:\mathcal{X}\to \mathbb{R}$ satisfying~\eqref{eq:data1}--\eqref{eq:data3} with $\gamma\le \delta<\eta$.
\end{enumerate}
\end{theorem}

\textit{Data-driven barrier certificates.} Theorem~\ref{thm:datamonbar} reduces safety verification over \(\Xx\), namely
conditions~\eqref{eq:indbar1-c}--\eqref{eq:indbar3-c}, to the finite
verification problem~\eqref{eq:data1-c}--\eqref{eq:data3-c}, evaluated at
partition extremal points.
Thus, the conditions can be checked using finitely many simulator queries. This
reduction exploits the monotone barrier structure of
Theorem~\ref{thm:monbar} and yields a scalable data-driven certification
framework.

\textit{Memory consumption.} For large \(K\), verifying conditions~\eqref{eq:data1-c}--\eqref{eq:data3-c}
can be memory-intensive, since it requires storing intermediate barrier
evaluations along trajectories, e.g.,
\(\mathcal{B}(f^{1}(x,\kappa(x))),\ldots,\mathcal{B}(f^{K}(x,\kappa(x)))\).
This overhead becomes significant for long horizons.
In Appendix~\ref{app:me}, we leverage barrier monotonicity to derive a
memory-efficient formulation of control, respectively robust, \(K\)-inductive
barrier certificates.

\section{Monotone neural barrier certificates}

Theorem~\ref{thm:datamonbar} addresses Problems~\ref{p2} and~\ref{p1} using
finitely many simulator queries, but does not provide a systematic way to find
monotone inductive barrier certificates \(\mathcal{B}\) satisfying
conditions~\eqref{eq:data1-c}--\eqref{eq:data3-c}.
In this section, we use \emph{monotone neural networks} to learn such
certificates via their approximation power.

\subsection{Monotone neural networks}\label{subsec:monNN}

Several monotone neural-network constructions have been proposed and applied
across domains~\cite{sill1997monotonic,liu2020certified,HD-MV:10,SY-DD-KC-JP-MG:17}.
A common architecture uses feedforward networks with nonnegative weights and
non-decreasing activations, guaranteeing monotonicity~\cite{HD-MV:10}. Despite
their simplicity, these networks can have limited expressive power. In
particular, with convex activations such as ReLU, the represented functions are
themselves convex~\citep[Lemma 1]{mikulincer2022size}.

Motivated by these limitations, we introduce \emph{Max-Monotone Networks},
a neural architecture with universal approximation guarantees for monotone
functions, tailored to learning inductive barrier certificates
satisfying~\eqref{eq:data1-c}--\eqref{eq:data3-c}. 
The construction is inspired by the fact that, under suitable assumptions, such
certificates admit a min-max-separable representation
\begin{align}\label{eq:min-max-separable}
\mathcal{B}(x)=\min_{j\in J}\{\max_{i\in\{1,\ldots,n\}} h_{i,j}(x_i)\}    
\end{align}
where each
\(h_{i,j}:\real\to\real\) is monotone. The proof is provided in Appendix~\ref{App:justminmax}.

Despite its simplicity, learning the min-max-separable representation~\eqref{eq:min-max-separable} is impractical in high
dimensions, as it requires a separate monotone one-dimensional network for each coordinate. A natural simplification is to share the component function,
yielding
$\mathcal{B}(x)=\max_{i\in\{1,\ldots,n\}} h(x_i) = h\left(\max_{i\in\{1,\ldots,n\}} x_i\right)$.
This motivates architectures built from coordinate-wise maximization features.
However, using only the feature \(\max_{i\in\{1,\ldots,n\}} x_i\) is too
restrictive. To increase expressiveness, we build on classical monotone
min--max networks~\cite{sill1997monotonic} by combining a coordinate-wise
maximum feature with multiple monotone feature maps obtained as maxima of
monotone affine transformations of the input.

\begin{definition}[Max-Monotone Networks]\label{eq:monnn}
\label{def:monNN}
A Max-Monotone Network is a mapping $N:\real^{n}\to\mathbb{R}$ of the form $N(x) = \min_{j\in\{0,\ldots,J\}} N^{(j)}\big(g^{(j)}(x)\big)$
where, for each $j\in\{0,\ldots,J\}$, $N^{(j)}:\real \to \mathbb{R}$ is a monotone neural network with nonnegative weights and a non-decreasing activation function $\sigma$, and the feature map $g^{(j)}:\real^n\to \real$ is defined by
\begin{align}
g^{(0)}(x) &= \max_{i\in\{1,\ldots,n\}} x_i,\quad
g^{(j)}(x) = \max_{k\in\{1,\ldots,m_j\}} { (W_j x + b_j)_k }, \quad j\in {1,\ldots,J},\label{eq:NNdef1}
\end{align}
with matrices $W_j \in \mathbb{R}^{m_j \times n}$ satisfying $W_j \ge 0$.
\end{definition}
Appendix~\ref{App:proofminmax} provides a schematic illustration of the
Max-Monotone Network architecture.
The next theorem establishes the monotonicity and universal approximation
properties of this architecture; the proof is referred to
Appendix~\ref{app:minmaxnnproof}.

\begin{theorem}[Max-Monotone Networks]\label{thm:minmaxnn}
A Max-Monotone Network is monotone. Moreover, for any continuous, bounded,
monotone \(m:[0,1]^n\to\mathbb{R}\) with bounded partial derivatives and any
\(\varepsilon>0\), there exists a Max-Monotone Network \(N\) such that
\(|m(x)-N(x)|<\varepsilon\), for all \(x\in[0,1]^n\).
\end{theorem}

\subsection{Training monotone neural barrier certificates}

Ensuring that a neural network satisfies
conditions~\eqref{eq:data1-c}--\eqref{eq:data3-c} or
\eqref{eq:data1}--\eqref{eq:data3} is challenging.
We therefore design loss functions for monotone neural networks that promote these constraints during training.

For safe control synthesis (Problem~\ref{p2}), we jointly train two monotone neural networks: a monotone neural barrier certificate \(N:\mathcal{X}\to\mathbb{R}\) parameterized as in Definition~\ref{eq:monnn}, and a monotone feedback policy \(\kappa:\mathcal{X}\to U\) with the same architecture except that each branch outputs a vector in \(\mathbb{R}^m\) and the final minimum is applied componentwise across branches.
Both networks are trained with the squared \(\Relu\)-error loss
\(L_{\text{total}}^{\text{s}} := L_1^{\text{s}} + L_2^{\text{s}} +
L_3^{\text{s}}\).
The terms \(L_1^{\text{s}}\), \(L_2^{\text{s}}\), and \(L_3^{\text{s}}\)
enforce the data-driven barrier constraints~\eqref{eq:data1-c},
\eqref{eq:data2-c}, and~\eqref{eq:data3-c}, respectively:
\begin{align}
L_1^{\text{s}} &= \sum_{i \in \mathcal{I}}[\Relu( \max_{k\in [K-1]} ( N(f^{k}(\bar{x}^i,\kappa(\bar{x}^i)) - \gamma )]^2, \label{eq:loss1_c}\\
L_2^{\text{s}} &= \sum_{i \in \mathcal{U}}[\Relu( \eta - N(\underline{x}^i)  )]^2, \label{eq:loss2_c}\\ 
L_3^{\text{s}} & = \sum_{i \in \mathcal{P}} r_{i} [\Relu( N(f^{K}(\bar{x}^i,\kappa(\bar{x}^i))) - \delta )]^2 \label{eq:loss3_c},
\end{align}
where $r_i
=
1-
\Relu\!\left(
\operatorname{sign}\!\left(
\max_{k\in[K-1]}
N\!\left(
f^{k}(\underline{x}^{i},\kappa(\underline{x}^{i}))
\right)
-\delta
\right)
\right)$, for every $i\in \mathcal{P}$. The parameters \(\gamma,\delta,\eta\in\mathbb{R}\), with
\(\gamma\le\delta<\eta\), are tunable hyperparameters that act as margins to
improve the robustness of the barrier certificate.
Training uses gradient-based optimization of \(L_{\text{total}}^{\text{s}}\)
and terminates once \(L_{\text{total}}^{\text{s}}=0\). 

For the robust safety verification problem (Problem~\ref{p1}), we train a monotone neural network
$N : \mathcal{X} \to \mathbb{R}$ as a monotone robust barrier certificate using a squared ReLU-error loss. The total loss is defined as
$L_{\text{total}}^{\text{v}} = L_1^{\text{v}} + L_2^{\text{v}} + L_3^{\text{v}}$,
where each term corresponds to one of the barrier certificate conditions.

The individual loss terms are given by
\begin{align}
L_1^{\text{v}} &= \sum_{i \in \mathcal{I},\bar{\mathbf{u}}\in \overline{\mathcal{U}}^K}[\Relu( \max_{k\in [K-1]} ( N(f^{k}(\bar{x}^i,\bar{\mathbf{u}})) - \gamma )]^2, \label{eq:loss1}\\
L_2^{\text{v}} &= \sum_{i \in \mathcal{U}}[\Relu( \eta - N(\underline{x}^i)  )]^2,\label{eq:loss2}\\ 
L_3^{\text{v}} & = \sum_{i \in \mathcal{P}, \bar{\mathbf{u}}\in \bar{\mathcal{U}}^{K+1}} r_{i} [\Relu( N(f^{K}(\bar{x}^i,\bar{\mathbf{u}})) - \delta )]^2,\label{eq:loss3}
\end{align}
where
\[
r_i
=
1-\Relu\!\left(
\operatorname{sign}\!\left(
\min_{\underline{\mathbf u}\in \underline{\mathcal U}^{K}}
\max_{k\in[K-1]}
N\!\left(f^k(\underline{x}^i,\underline{\mathbf u})\right)
-\delta
\right)
\right).
\]

Here, $\operatorname{sign}(0)=0$ and $\gamma\le\delta<\eta$ are the certificate thresholds.

\textit{Safety certificates from zero loss.} Achieving zero loss, \(L^{\text{s}}_{\text{total}}=0\) (resp. \(L^{\text{v}}_{\text{total}}=0\)), is equivalent to the neural network $N$ satisfying the
inductive barrier conditions~\eqref{eq:data1-c}--\eqref{eq:data3-c} (resp. conditions~\eqref{eq:data1}--\eqref{eq:data3}).
Thus, the learned monotone neural network \(N\) is an inductive barrier
certificate that formally certifies system safety. In the next theorem, we show that the loss functions defined in~\eqref{eq:loss1_c}--\eqref{eq:loss3_c} (resp. defined in~\eqref{eq:loss1}--\eqref{eq:loss3}) not only promote the learned monotone neural network to satisfy conditions~\eqref{eq:data1-c}--\eqref{eq:data3-c} (resp. conditions~\eqref{eq:data1}--\eqref{eq:data3}), but also guarantee that, whenever these loss functions vanish, the resulting network constitutes an inductive barrier certificate for $\Sys$. We refer to Appendix~\ref{App:proofdata} for the proof.

\begin{theorem}[Zero-loss certification]
\label{thm:zeroloss}
Consider a monotone dtDS $\mathfrak{S} = (\mathcal{X}, \mathcal{X}_0, U, f)$ with an unknown state-transition map $f$ and a set of unsafe states $\mathcal{X}_u$. Let $\{[\underline{x}^i, \bar{x}^i]\}_{i \in \mathcal{P}}$ be a family of hyper-rectangles covering the state space $\mathcal{X}$, and let $\{[\underline{u}^j, \bar{u}^j]\}_{j \in \mathcal{Q}}$ be a family of hyper-rectangles covering the input space $U$. Define the index sets $\mathcal{I}$ and $\mathcal{U}$ as the minimal subsets of $\mathcal{P}$ satisfying \eqref{eq:cover}. Assume $\gamma<\delta<\eta$. Then,

\begin{enumerate} [label=\arabic*)]

\item\label{thm:zeroloss1} \textit{(Safe control synthesis)} If there exist monotone neural networks
$N:\mathcal X\to\mathbb R$ and $\kappa:\mathcal X\to U$
such that the loss in \eqref{eq:loss1_c}--\eqref{eq:loss3_c} satisfies
$L_{\mathrm{total}}^{\mathrm s}=0$,
then the system $\Sys$ is safely controllable with respect to $\mathcal X_u$ with the feedback controller $\kappa$.

\item\label{thm:zeroloss2} \textit{(Robust safety verification)} If there exists a monotone neural network $N:\mathcal X\to\mathbb R$
such that the loss in \eqref{eq:loss1}--\eqref{eq:loss3} satisfies
$L_{\mathrm{total}}^{\mathrm v}=0$,
then $\Sys$ is robustly safe with respect to $\mathcal X_u$.

\end{enumerate}
\end{theorem}

\textit{Neural network loss functions.}
The total loss \(L_{\text{total}}=L_1+L_2+L_3\) can be adjusted to improve
training, by replacing nonsmooth terms such as \(\Relu\) or
\(\operatorname{sign}\) with smoother MSE objectives, while using the original
barrier loss only for termination.
In our experiments, however, we directly optimize the proposed loss and use it
for termination, observing stable training and strong performance.

\noindent\textbf{Iterative refinement of the hyper-rectangular cover.} While zero training loss yields a formal safety certificate, convergence to zero
is not guaranteed and may be limited by insufficient samples. This motivates
adaptive sampling algorithms that iteratively refine the
partition of the state space, enriching the dataset with informative samples and
improving the chance of certification.
Appendix~\ref{App:grid} presents two refinement strategies: \emph{monotone
shrinking refinement} (Algorithm~\ref{alg:monotoneshrinking}) and
\emph{arbitrary refinement} (Algorithms~\ref{alg:arbgrid}--\ref{alg:inputgrid}).
Algorithm~\ref{alg:learningwithshrinking} (resp. Algorithm~\ref{alg:learningrobustwithshrinking}) integrates these refinements into
training Max-Monotone Networks with the losses~\eqref{eq:loss1_c}--\eqref{eq:loss3_c} (resp. losses~\eqref{eq:loss1}--\eqref{eq:loss3}).
Next, we establish soundness and simulator-query complexity bounds for Algorithm~\ref{alg:learningwithshrinking}; see Appendix~\ref{App:proof_shrinkcontrol} for the
proof.

\begin{theorem}[Soundness and linear simulator-query complexity of Algorithm~\ref{alg:learningwithshrinking}]
\label{thm:shrinkcontrolbarrier}
Consider a monotone dtDS $\mathfrak{S} = (\mathcal{X}, \mathcal{X}_0, U, f)$ with a set of unsafe states $\mathcal{X}_u\subseteq \Xx\subseteq \real^n$. Let $\{[\underline{x}^i, \bar{x}^i]\}_{i \in \mathcal{P}}$ be a family of hyper-rectangles covering $\mathcal{X}$, and let $\{[\underline{u}^j, \bar{u}^j]\}_{j \in \mathcal{Q}}$ be a family of hyper-rectangles covering $U$. Define the index sets $\mathcal{I}$ and $\mathcal{U}$ as the minimal subsets of $\mathcal{P}$ satisfying \eqref{eq:cover} and suppose that the cardinality of $\mathcal{P},\mathcal{I},\mathcal{U}$ are independent of $n$.
Then:
\begin{enumerate}[label=\arabic*)]
    \item \textit{(Soundness)} If Algorithm~\ref{alg:learningwithshrinking} returns $(N,\kappa,L_{\mathrm{total}})$, then the system $\Sys$ is safely controllable with respect to $\Xx_u$ using the controller $\kappa$.
    \item \textit{(Simulator-query complexity)} If Algorithm~\ref{alg:monotoneshrinking} is used as the refinement routine in line~10 of Algorithm~\ref{alg:learningwithshrinking}, then the total number of simulator queries required by Algorithm~\ref{alg:learningwithshrinking} is $\mathcal{O}(n)$.
\end{enumerate}
\end{theorem}

For high-dimensional systems, Algorithm~\ref{alg:learningwithshrinking} may
require many samples to construct a valid monotone barrier certificate.
The inductive parameter \(K\) helps mitigate this issue.
Rather than relying only on finer state-space partitions, larger \(K\) values
incorporate multi-step system evolution into the certification conditions,
reducing the partition resolution needed for certification and lowering sample
complexity.
Appendix~\ref{App:k} demonstrates the practical benefits of tuning \(K\) under a
fixed partition.

\begin{algorithm}[!h]
\small
\caption{Monotone neural control barrier learning with adaptive refinement}
\label{alg:learningwithshrinking}

\textbf{Inputs:} Simulator access to the dynamics $f$; hyper-rectangular covers
$\{[\underline{x}^{i},\bar{x}^{i}]\}_{i\in\mathcal P}$, $\{[\underline{x}_0^i,\bar{x}_0^i]\}_{i\in\mathcal I}$, and
$\{[\underline{x}_u^j,\bar{x}_u^j]\}_{j\in\mathcal U}$,
of $\Xx$, $\Xx_0$, and $\Xx_u$, respectively;
margins $\gamma<\delta<\eta$; maximum inductive parameter $K_{\max}$; maximum shrinking budget $n_{\max}^{\mathcal P}$; and maximum number of epochs $N_{\mathrm{ep}}$; a fixed per-coordinate search budget $N_{\mathrm{search}}\in\mathbb{N}_{>0}$; and constants $\epsilon>0$ and $C_{\max}>1+\epsilon$ for arbitrary refinement.

\textbf{Output:} A monotone neural control barrier certificate $N$, controller $\kappa$, and final loss $L_{\mathrm{total}}$, if found; otherwise \texttt{FAIL} and final loss $L_{\mathrm{total}}$.

\begin{algorithmic}[1]

\State Initialize
$\mathcal H \gets \{[\underline{x}^{\,\ell},\bar{x}^{\,\ell}]\}_{\ell\in\mathcal P\setminus\mathcal{U}},
\;\;
\mathcal C \gets \emptyset$,\;\; $L_{\mathrm{total}}\gets \infty$.

\For{$p=1,\ldots,n_{\max}^{\mathcal P}$}

    \For{$K=1,\ldots,K_{\max}$}

        \State Define
        $L_{\mathrm{shrink}}
        =
        \sum_{x^c\in\mathcal C}
        \big[\Relu(\eta-N(x^c))\big]^2$.

        \State Train $N$ and $\kappa$ for at most $N_{\mathrm{ep}}$ epochs using
        $L_{\mathrm{total}}
        =
        L_1+L_2+L_3+L_{\mathrm{shrink}}$,
        where 
        \Statex \hspace{\algorithmicindent} \hspace{\algorithmicindent}$L_1,L_2,L_3$ are defined in~\eqref{eq:loss1_c}--\eqref{eq:loss3_c}
        using $\mathcal I$, $\mathcal U$, the current unresolved index set induced         
        \Statex \hspace{\algorithmicindent} \hspace{\algorithmicindent}by $\mathcal H$, the current value of $K$, and margins $\gamma,\delta,\eta$.

        \If{$L_{\mathrm{total}}=0$}
            \State \textbf{return} $N,\kappa,L_{\mathrm{total}}$
        \EndIf

    \EndFor

    \State Apply either Algorithm~\ref{alg:monotoneshrinking}
    with budget $N_{\mathrm{search}}$ or
    Algorithm~\ref{alg:arbgrid} with parameters $\epsilon,C_{\max}$ to obtain $(\mathcal H^{+},\mathcal C^{+})$

    \If{the chosen refinement algorithm returns \texttt{FAIL}}
        \State \textbf{return} \texttt{FAIL}, $L_{\mathrm{total}}$
    \Else
        \State Update $(\mathcal H,\mathcal C)$ using the returned $(\mathcal H^{+},\mathcal C^{+})$
    \EndIf

\EndFor

\State \textbf{return} \texttt{FAIL}, $L_{\mathrm{total}}$

\end{algorithmic}
\end{algorithm}

\section{Numerical experiments}\label{sec:numericalexp}
In this section, we demonstrate the computational efficiency and scalability of the proposed framework across several representative benchmark systems. The dynamics are
provided for clarity but are not used during learning; each system is treated as
a black box, using only sampled data.
We compare with two baselines: the counterexample-guided approach
of~\cite{edwards2024fossil} and the Lipschitz-based method
of~\cite{anand2023formally}. Additional implementation details, hardware
specifications, neural architectures, and baseline configurations are given in
Appendix~\ref{App:models}.
Table~\ref{tb:trf-sc} shows scalability on Lotka--Volterra systems from 2 to
100 dimensions, while Table~\ref{tb:comp} compares runtime and sample complexity
across case studies. The results show that competing methods either fail to
converge or require exponentially many samples as dimension grows, whereas our
approach remains scalable and sample-efficient in high dimensions.

\begin{table}[t]
\centering
\small
\caption{Safety-verification performance on Lotka--Volterra with increasing dimension.
Averaged over 10 runs. NC: no convergence within 12 hours; EM: out of memory.
Our method uses \(K=2\).}
\vspace{4pt}
\label{tb:trf-sc}
\setlength{\tabcolsep}{10pt}

\begin{tabular}{c cc @{\hspace{35pt}} cc @{\hspace{35pt}}cc}
\toprule
\multirow{2}{*}{\textbf{Dim}} 
& \multicolumn{2}{c}{\textbf{\cite{edwards2024fossil}}} {\hspace{35pt}}
& \multicolumn{2}{c}{\textbf{\cite{anand2023formally}}} {\hspace{35pt}}
& \multicolumn{2}{c}{\textbf{Ours}} \\
& Time & Samples & Time & Samples & Time & Samples \\
\midrule
2   & 47.5s  & 1000 & 85.9s  & $10^6$           & \textbf{5.2s}   & \textbf{9} \\
3   & NC     & 1000 & NC     & $2^6\times10^{6}$ & \textbf{6.2s}   & \textbf{10} \\
5   & NC     & 1000 & NC     & $2^6\times10^{6}$ & \textbf{5.8s}   & \textbf{12} \\
10  & NC     & 1000 & NC     & $2^6\times10^{6}$ & \textbf{5.9s}   & \textbf{17} \\
50  & NC     & 1000 & EM     & $\Omega(2^{50})$  & \textbf{6.6s}   & \textbf{57} \\
100 & NC     & 1000 & EM     & $\Omega(2^{100})$ & \textbf{10.1s}  & \textbf{107} \\
\bottomrule
\end{tabular}
\end{table}

\noindent\textbf{Freeway traffic flow dynamics.} We consider the Daganzo cell transmission model (CTM)~\cite{CFD:94}, which
captures demand and supply constraints on a line-digraph traffic network
(Fig.~\ref{fig:celltran}). This model induces a monotone system
\(\mathfrak{S}=(\Xx,\Xx_0,U,f)\)~\citep[Lemma 4.28]{FB-CTDS}, with
\(\Xx=[0,50]^n\), \(\Xx_0=[0,5]^n\), and \(U=[0,7.5]\). The unsafe set is
\(\Xx_u=[0,50]^n\setminus[0,40]^n\).
Our framework certifies safety for \(n=1000\), \(K=50000\), using \(151{,}001\)
samples and trains a monotone neural robust barrier certificate in \(542.2\)~s
on average over \(10\) runs. Details are in Appendix~\ref{App:traffic}.

\noindent\textbf{Network of coupled oscillators.}
We consider discrete-time Kuramoto oscillator networks~\cite{skardal2015control}
with all-to-all topology and with topologies from IEEE power-system testbeds from \textsc{Matpower} black-box model~\cite{matpower2011}, yielding
monotone systems \(\mathfrak{S}=(\Xx,\Xx_0,U,f)\). The state space is
\(\Xx=[0,\tfrac{\pi}{4}]^n\), the initial set is
\(\Xx_0=[0,\tfrac{\pi}{16}]^n\), and the input set is
\(U=[-\tfrac{\pi}{4},\tfrac{\pi}{4}]^n\). The unsafe set is
\(\Xx_u=[0,\tfrac{\pi}{4}]^n\setminus[0,\tfrac{\pi}{8}]^n\).
For all-to-all graph structure, we certify safety for \(n=9000\), \(K=2\), using
\(9{,}007\) samples and synthesizing a safe control certificate in \(395.6\)~s
on average over \(10\) runs. For IEEE-based topologies, we consider IEEE-118
(\(n=118\)), IEEE-2383WP (\(n=2383\)), and IEEE-13659-PEGASE (\(n=13659\)),
synthesizing safe control barrier certificates in \(9.7\), \(30.4\), and
\(787.1\)~s, respectively, with \(K=2\). Further details are in
Appendix~\ref{App:kuramoto}.

\begin{table}[!t]
\centering
\small
\caption{Runtime and sample complexity across case studies (averaged over 10 runs). NC indicates failure to converge within 12 hours, and EM indicates exceeding available memory.}
\label{tb:comp}
\vspace{4pt}

\renewcommand{\arraystretch}{1.2}
\setlength{\tabcolsep}{8pt}

\begin{tabular}{l ccc @{\hspace{40pt}} ccc}
\toprule
\multirow{2}{*}{\textbf{Case (Dim)}} 
& \multicolumn{3}{c}{\textbf{Time (s)}} {\hspace{40pt}}
& \multicolumn{3}{c}{\textbf{Samples}} \\
& \cite{edwards2024fossil} & \cite{anand2023formally} & Ours & \cite{edwards2024fossil} & \cite{anand2023formally} & Ours \\
\midrule
Freeway traffic (1{,}000D)      & NC & EM & \textbf{542.2} & $10^3$ & $\Omega(2^{1000})$ & \textbf{151{,}001} \\
Kuramoto (all-to-all) (9{,}000D)     & NC & EM & \textbf{395.6} & $10^3$ & $\Omega(2^{9000})$ & \textbf{9{,}007} \\
IEEE 118-Bus (118D) & NC & EM & \textbf{9.7}  & $10^3$ & $\Omega(2^{118})$ & \textbf{125} \\
IEEE 2383WP (2{,}383D) & NC & EM & \textbf{30.4}  & $10^3$ & $\Omega(2^{2383})$ & \textbf{2{,}390} \\
PEGASE (13{,}659D) & NC & EM & \textbf{787.1}  & $10^3$ & $\Omega(2^{13659})$ & \textbf{13{,}666} \\
Lotka--Volterra (2{,}000D) & NC & EM & \textbf{62.3}  & $10^3$ & $\Omega(2^{2000})$ & \textbf{2{,}301} \\
Ribosome Flow Model (500D) & NC & EM & \textbf{3712.1}  & $10^3$ & $\Omega(2^{500})$ & \textbf{8{,}001} \\
Ribosome Flow Model (1{,}000D) & NC & EM & \textbf{14112.5}  & $10^3$ & $\Omega(2^{1000})$ & \textbf{16{,}001} \\
\bottomrule
\end{tabular}
\end{table}

\noindent\textbf{Population dynamics.}
We consider a discrete-time Lotka--Volterra population model~\cite{FB-CTDS},
which induces a monotone system \(\mathfrak{S}=(\Xx,\Xx_0,f)\) on the state space \(\Xx=[0,1]^n\) under the parameter and step-size conditions given in Appendix~\ref{App:lvmon}. The initial set is
\(\Xx_0=[0,0.4]^n\), and the unsafe set is
\(\Xx_u=[0,1]^n\setminus[0,0.5]^n\).
Our framework certifies safety for \(n=2000\), \(K=100\), using \(2{,}301\)
samples and learning a monotone neural robust barrier certificate in \(62.3\)~s
on average over \(10\) runs. Further details are in Appendix~\ref{App:lvmon}.

\noindent\textbf{Ribosome Flow Model.}
We consider a Ribosome Flow Model (RFM)~\cite{MA-DA-EDS:20,MM-TT:12} with
\(n\) sites, inducing a monotone system \(\mathfrak{S}=(\Xx,\Xx_0,f)\). The
state space is \(\Xx=[0,1]^n\), the initial set is \(\Xx_0=[0,0.2]^n\), and
the unsafe set is \(\Xx_u=[0,1]^n\setminus[0,0.9]^n\).
Our framework certifies safety for \(n=500\), \(K=2500\), using \(8{,}001\)
samples and learning a valid barrier certificate in \(3712.1\)~s over \(10\)
runs. For \(n=1000\), it certifies safety with \(K=5000\), using \(16{,}001\)
samples in \(14122.5\)~s. Details are in Appendix~\ref{App:rfm}.

\subsection{Ablation Studies}

We conduct two ablation studies to evaluate the contributions of the Max-Monotone architecture and adaptive refinement in learning monotone $K$-inductive barrier certificates.

\paragraph{Network architecture.}
We compare Max-Monotone Networks against a nonnegative-weight MLP~\cite{HD-MV:10} and a classical monotone min--max network~\cite{sill1997monotonic} for learning monotone $K$-inductive barrier certificates. The certificate conditions, data, optimizer, initialization protocol, refinement procedure, and 12-hour computational budget are fixed, with parameter counts matched exactly for the MLP and within 7\% for the min--max network.
As shown in Table~\ref{tab:arch_ablation}, only Max-Monotone Network achieves certification within the prescribed budget, demonstrating its effectiveness for certificate learning without implying that the alternative architectures lack
sufficient expressiveness.

\begin{table}[!ht]
\centering
\small
\caption{Architecture ablation under a 12-hour computational budget.}
\label{tab:arch_ablation}
\begin{tabular}{lcc}
\toprule
Architecture & Lotka--Volterra (2,000D) & Traffic (1,000D) \\
\midrule
Nonnegative-weight MLP & Not certified & Not certified \\
Classical monotone min--max & Not certified & Not certified \\
\textbf{Max-Monotone (ours)} &
\textbf{62.3 s; 2,301 samples} &
\textbf{542.2 s; 151,001 samples} \\
\bottomrule
\end{tabular}
\end{table}

\paragraph{Adaptive refinement.}
We also examine the effect of adaptive refinement on the 2D traffic example with $K=1$, keeping the network architecture and training settings fixed. As shown in Table~\ref{tab:ref_ablation}, the original four-box partition fails to certify safety, whereas adaptive refinement achieves certification using 103,000 final boxes. This highlights the role of refinement in tightening finite barrier conditions.

\begin{table}[!ht]
\centering
\small
\caption{Adaptive-refinement ablation on the 2D traffic example ($K=1$).}
\label{tab:ref_ablation}
\begin{tabular}{lcc}
\toprule
Refinement & Final boxes & Outcome \\
\midrule
Disabled & 4 & Not certified \\
Enabled & 103,000 & \textbf{Certified} \\
\bottomrule
\end{tabular}
\end{table}

\section{Conclusion}
This paper introduced a scalable, learning-based framework for synthesizing
safety controllers for high-dimensional unknown \emph{monotone} systems directly from
data. 
We showed that barrier certificates for monotone systems inherit the underlying monotonicity structure of the dynamics, thereby reducing safety verification to a finite collection of localized boundary checks. Building on this structural property, we proposed a specialized neural network architecture, called \emph{Max-Monotone Networks}, for learning such barrier certificates directly from data. Our approach yields substantially improved sample efficiency over
existing black-box methods and can avoid exponential growth in sample
complexity. Experiments on large-scale benchmarks further show that our method scales well in both data usage and training stability, even in high-dimensional settings. 

\begin{ack}
The authors thank the anonymous reviewers for their insightful feedback. This work was supported in part by NSF CAREER grant CNS-2145184 and NSF grants CNS-2111688 and CNS-2534982. The authors declare no competing interests.
\end{ack}

\bibliographystyle{abbrv}
\bibliography{ref,SJ,ref2}

\newpage
\appendix

\section{Related Works}\label{app:relatedwork}

\paragraph{Barrier Certificates.} The work \cite{prajna2004safety} introduced barrier certificates (BCs) as functions whose level sets over-approximate system reachable sets. Control barrier certificates (CBCs) later emerged as a key tool for synthesizing safe controllers~\citep{xiao2019control,jagtap2020formal,clark2021verification,dai2023convex}, typically using Sum-of-Squares (SOS) optimization~\citep{prajna2007framework,zhao2023safety,schneeberger2023sos}.

\paragraph{Neural Barrier Certificates.} The paper \cite{zhao2020synthesizing} introduced neural network–based barrier certificates, using a shallow network and MILP for formal verification. Later, \cite{peruffo2021automated} employed SMT solvers to find counterexamples for iterative refinement. Neural barrier certificates have since been extended to hybrid systems~\citep{zhao2021synthesizing} and stochastic dynamics~\citep{mathiesen2022safety}.

\paragraph{Neural Control Barrier Certificates.} Neural control barrier certificates (NCBCs) extend these ideas to safe controller synthesis. Recent work~\citep{robey2020learning,lindemann2021learning,dawson2022safe,dawson2023safe,liu2023safe,lindemann2024learning} learns both barrier functions and control policies using neural networks. Verification typically relies on SMT solvers~\citep{abate2020formal,zhao2020synthesizing,edwards2024fossil}, reachable set methods~\citep{xiang2018output}, polynomial approximations~\citep{sha2021synthesizing}, Lipschitz bounds~\citep{anand2023formally}, or ReLU verification~\citep{katz2017reluplex}. Recently,~\citep{zhang2024exact,zhang2024seev} an exact NCBC verification algorithm was proposed that synthesizes a controller if the barrier is valid. Most of these methods, however, assume full system knowledge (except~\citep{anand2023formally}) and rely heavily on SMT solvers, which scale poorly to deep networks due to their exponential complexity.

\section{Proof of Theorem~\ref{thm:monbar}}\label{App:proofiff}

We start by proving Theorem~\ref{thm:monbar}~part~\ref{p2}. We divide the proof into \textbf{if} part and \textbf{only if} part.

\paragraph{Proof of \textit{if part}.} Suppose that $B:\Xx\to \real$ is a monotone barrier certificate satisfying~\eqref{eq:indbar1-c},~\eqref{eq:indbar2-c}, and~\eqref{eq:indbar3-c}. Then, by condition~\eqref{eq:indbar1-c}, for any $x_0\in \Xx_0$, there exists $\mathbf{u} = (u_0,\ldots,u_{K-2})\in U^{K-1}$ such that $B(f^{k}(x_0,\mathbf{u})) \le \gamma$, for every $k\in [K-1]$. Since $\gamma\le \delta$, this means that $B(f^{k}(x_0,\mathbf{u})) \le \delta$, for every $k\in [K-1]$. Using condition~\eqref{eq:indbar3-c}, we deduce that there exists $u_{K-1}\in U$ such that 
\begin{align*}
    {B}(f(f^{K-1}(x_0,\mathbf{u}),u_{K-1}))\le \delta. 
\end{align*}
Now consider the point $x_1=f(x_0,u_0)\in \mathcal{X}$ and the control sequence $\widehat{\mathbf{u}}=(u_{1},\ldots,u_{K-1})\in U^{K-1}$. It is clear that, $\mathbb{B}(f^{k}(x_1,\widehat{\mathbf{u}}))= \mathbb{B}(f^{k+1}(x_0,\mathbf{u}^+))$, for every $k\in [K-1]$, where $\mathbf{u}^{+} = (u_0,\ldots,u_{K-1})\in U^K$. Therefore, for every $k\in [K-1]$, 
\begin{align*}
    {B}(f^{k}(x_1,\widehat{\mathbf{u}}))\le \delta.
\end{align*}
Using condition~\eqref{eq:indbar3-c}, we deduce that there exists $u_{K}\in U$ such that 
\begin{align*}
    {B}(f(f^{K}(x_0,\mathbf{u}^{+}),u_{K})) = \mathbb{B}(f(f^{K-1}(x_1,\widehat{\mathbf{u}}),u_{K}))\le \delta. 
\end{align*}
Thus, by iteratively applying this procedure, we can find a sequence $\{u_t\}_{t=0}^{\infty}$ such that, for every $t\ge 0$, we have 
\begin{align*}
    {B}(x_{t+1}) \le \delta. 
\end{align*}
where $x_{t+1} = f(x_t,u_t)$, for every $t\ge 0$. This implies that $\{x_t\}_{t=0}^{\infty} \subseteq \{y\in \real^n\mid \mathbb{B}(y)\le \delta\}$. On the other hand, by condition~\eqref{eq:indbar2-c}, we know that, for every $x\in \Xx_u$, we have $B(x)\ge \eta > \delta$. As a result $\{x_t\}_{t=0}^{\infty}\cap \Xx_u=\emptyset$. This means that the system $\Sys$ is safe with the feedback controller $\kappa:\Xx\to \real$.

\paragraph{Proof of \textit{only if part}.} Suppose that the system $\Sys$ is safely controllable with respect to $\Xx_u$. Before we state the proof, we introduce several notations and intermediate results that will be used throughout the proof. Since $\Sys$ is safely controllable with respect to $\Xx_u$, for every $x_0\in \Xx_0$, there exists a control sequence $\{u_t\}_{t=0}^{\infty}$ such that the associated trajectory $\{x_t\}_{t=0}^{\infty}$ avoids the unsafe set $\Xx_u$. We label this specific control sequence associated to $x_0\in \Xx_0$ by $\{u^{x_0}_t\}_{t=0}^{\infty}$. The \textit{reachable set} of $\Sys$, which is denoted by $\Omega\subseteq \Xx$, is defined by 
\begin{equation}\label{eq:reachsetk}
\begin{aligned}
\Omega &= \bigcup_{T\in \mathbb{Z}_{\ge 0}}\Omega_T,\\
    \Omega_T &=\setdef{x_T\in \Xx}{\{x_t\}_{t=0}^{\infty}\mbox{ is a traj of $\Sys$ starting from }x_0\in \Xx_0 \mbox{ with controller } \{u^{x_0}_t\}_{t=0}^{\infty}}.
\end{aligned}
\end{equation}
Motivated by the definition of the reachable set of $\Sys$ in~\eqref{eq:reachsetk} we define the set $\widehat{\Omega}\subseteq \Xx$ as the smallest lower closed set in $\Xx$ containing $\Omega$. More specifically, $\widehat{\Omega}$ is defined as:
\begin{equation}\label{eq:reachset-lowerk}
\begin{aligned}
\widehat{\Omega} =\setdef{x\in \Xx}{x\le y, \mbox{ for some }y\in \Omega}.
\end{aligned}
\end{equation}
Given $\gamma,\eta\in \real_{\ge 0}$ such that $\gamma\le \eta$, we use the lower closed set $\widehat{\Omega}$ to define the function $B:\Xx\to \real$ as follows:
\begin{align}\label{eq:Bk}
B(x) = \begin{cases}\gamma & x\in \widehat{\Omega}\\
\eta+1 & x\not\in \widehat{\Omega}\end{cases}    
\end{align}
The next lemma proves a useful property of the function $B$. 
\begin{lemma}\label{lem:Bmonk}
Consider a monotone dtDS $\Sys=(\Xx,\Xx_0,U,f)$ with the lower closed set $\widehat{\Omega}$ as defined in~\eqref{eq:reachset-lower}. Then the function $B:\Xx\to \real$ defined in~\eqref{eq:Bk} is monotone.
\end{lemma}
\begin{proof}
    Let $x,y\in \Xx$ be such that $x\le y$. We consider two cases: i) $y\in \widehat{\Omega}$, and ii) $y\not\in \widehat{\Omega}$. 
    \begin{enumerate}
        \item In this case, there exists $z\in \Omega$ such that $y\le z$. Note that we have $x\le y\le z$. This implies that $x\in \widehat{\Omega}$. As a result, by definition of $B$ in~\eqref{eq:Bk}. we have $B(x)=B(y)=\gamma$.
        \item In this case, by definition of $B$ in~\eqref{eq:Bk}, we have $B(y) = \eta+1$. This means that $B(x)\le \eta+1 = B(y)$.
    \end{enumerate}
    In both cases i) and ii), we have $B(x)\le B(y)$. This implies that $B$ is a monotone map.
\end{proof}

Now we get back to Theorem~\ref{thm:monbar} part~\ref{p2}.

Let the lower closed set $\widehat{\Omega}$ be defined in~\eqref{eq:reachset-lowerk} and the function $B:\Xx\to \real$ be defined as in~\eqref{eq:Bk}. Our goal is to show that $B:\Xx\to \real$ is a monotone $K$-inductive control barrier certificate for $\Sys$ satisfying conditions~\eqref{eq:indbar1-c},\eqref{eq:indbar2-c}, and~\eqref{eq:indbar3-c}. By Lemma~\ref{lem:Bmonk}, it is clear that $B$ is monotone. 

 \begin{enumerate}
    \item\textbf{($B$ satisfies condition~\eqref{eq:indbar1-c})} Suppose that $x_0\in \Xx_0$. Since $\Sys$ is safely controllable with respect to $\Xx_u$, there exists a controller $\{u^{x_0}_t\}_{t=0}^{\infty}$ such that system $\Sys$ under this controller starting from $x_0$ never enters $\Xx_u$. Let $\mathbf{u}=(u^{x_0}_0,u^{x_0}_1,\ldots,u^{x_0}_{K-1})$. By the definition of $\Omega$, this implies that $x_0 \in \Omega$. Moreover, $f^{k}(x_0,\mathbf{u}) \in \Omega$, for every $k\in [K-1]$. As a result, we have $f^{k}(x_0,\mathbf{u})\in \Omega \subseteq \widehat{\Omega}$, for every $k\in [K-1]$. Using the definition of $B$ in~\eqref{eq:Bk}, we get $B(f^{k}(x_0,\mathbf{u})) \le \gamma$, for every $k\in [K-1]$. This shows that $B$ satisfies condition~\eqref{eq:indbar1-c} with the control input $\mathbf{u}$.
    
    \item\textbf{($B$ satisfies condition~\eqref{eq:indbar2-c})} Suppose that $x\in \Xx_u$. Since $\Sys$ is safe with respect to $\Xx_u$, we get that $x\not\in \Omega$. Note that $\widehat{\Omega}$ is the smallest lower closed set containing $\Omega$ and $\Xx_u$ is an upper closed set. As a result, $x\not\in \widehat{\Omega}$. By definition of $B$ in~\eqref{eq:Bk}, we have $B(x) =\eta +1 > \gamma$. This shows that $B$ satisfies condition~\eqref{eq:indbar2-c}.
    
    \item\textbf{($B$ satisfies condition~\eqref{eq:indbar3-c})} Let $x\in \Xx$ and the control input $\mathbf{u} = (u_0,\ldots,u_{K-2})\in U^{K-1}$ is such that $B(f^{k}(x,\mathbf{u}))\le \delta$, for every $k\in [K-1]$. By definition of $B$ and using the fact that $\delta<\eta$, we have $f^{k}(x,\mathbf{u})\in \widehat{\Omega}$, for every $k\in [K-1]$. This means that, there exists $y\in \Omega$ such that $f^{K-1}(x,\mathbf{u})\le y$. This implies that, there exists $T>0$, a point $x_0\in \Xx_0$ and an associated controller $\{u^{x_0}_t\}_{t=0}^{\infty}$ such that $y = x_T$, where $x_{t+1}=f(x_t,u^{x_0}_t)$, for every $t\ge 0$. Since $f$ is monotone, we get
    \begin{align*}
        f(f^{K-1}(x,\mathbf{u}),u^{x_0}_T)\le f(y,u^{x_0}_T) = f(x_T,u^{x_0}_T) = x_{T+1}\in \Omega_{T+1}\subseteq \Omega\subseteq \widehat{\Omega}. 
        \end{align*}
        Therefore $B(f(f^{K-1}(x,\mathbf{u}),u^{x_0}_T) = \gamma \le \delta$.  This shows that $B$ satisfies condition~\eqref{eq:indbar3-c}.
\end{enumerate}
This completes the proof of Theorem~\ref{thm:monbar}~part~\ref{p2}.

Now we focus on Theorem~\ref{thm:monbar} part~\ref{p1}. Before proving Theorem~\ref{thm:monbar} part~\ref{p1}, we introduce several notations and intermediate results that will be used throughout the proof. Consider a monotone dtDS $\Sys=(\Xx,\Xx_0,U,f)$. The \textit{reachable set} of $\Sys$, denoted by $\Omega\subseteq \Xx$, is defined as follows: 
\begin{equation}\label{eq:reachset}
\begin{aligned}
\Omega &= \bigcup_{T\in \mathbb{Z}_{\ge 0}}\Omega_T,\\
    \Omega_T &=\setdef{x_T\in \Xx}{\{x_t\}_{t=0}^{\infty}\mbox{ is a traj of $\Sys$ with some input }\{u_t\}_{t=0}^{\infty} \mbox{ with } x_0\in \Xx_0}.
\end{aligned}
\end{equation}
We first state the following intermediate result about the reachable set $\Omega$. 
\begin{lemma}\label{lem:reach}
   Consider a monotone dtDS $\Sys=(\Xx,\Xx_0,U,f)$ with the reachable set $\Omega$ defined in~\eqref{eq:reachset}. Then, for every $x\in \Omega$ and every $u\in U$, we have $f(x,u)\in \Omega$. 
\end{lemma}
\begin{proof}
    Let $x\in \Omega$ and $u\in U$. Then there exists $T\in \mathbb{Z}_{\ge 0}$ such that $x\in \Omega_{T}$. This means that there exists $\{x_t\}_{t=0}^{\infty}$ a trajectory of $\Sys$ with the input $\{u_t\}_{t=0}^{\infty}$ starting from $x_0\in \Xx_0$ such that $x=x_T$. Now consider the input $\{w_t\}_{t=0}^{\infty}$ defined by
    \begin{align}\label{eq:w}
        w_t = \begin{cases}
            u_t & t < T,\\
            u & t \ge T.
        \end{cases}
    \end{align}
    Let $\{y_t\}_{t=0}^{\infty}$ be the trajectory of $\Sys$ with the input $\{w_t\}_{t=0}^{\infty}$ starting from $x_0$. By definition of the input $\{w_t\}_{t=0}^{\infty}$ in~\eqref{eq:w}, it is clear that $y_t = x_t$, for every $t\le T$. Moreover, for $t = T+1$, we have 
    \begin{align*}
        y_{T+1} = f(y_T,w_T) = f(y_T,u) = f(x_T,u) = f(x,u), 
    \end{align*}
    where the second equality holds by definition of the input $\{w_t\}_{t=0}^{\infty}$ in~\eqref{eq:w}, the third equality holds using the fact that $x_t = y_t$, for every $t\le T$, and the fourth equality holds because $x=x_T$. Therefore, by definition of $\Omega_{T+1}$, we have $f(x,u) \in \Omega_{T+1}$ and thus  $f(x,u)\in \Omega$. 
\end{proof}
Motivated by the definition of the reachable set of $\Sys$ in~\eqref{eq:reachset} we define the set $\widehat{\Omega}\subseteq \Xx$ as the smallest lower closed set in $\Xx$ containing $\Omega$. More specifically, $\widehat{\Omega}$ is defined as:
\begin{equation}\label{eq:reachset-lower}
\begin{aligned}
\widehat{\Omega} =\setdef{x\in \Xx}{x\le y, \mbox{ for some }y\in \Omega}.
\end{aligned}
\end{equation}
Given $\gamma,\eta\in \real_{\ge 0}$ such that $\gamma\le \eta$, we use the lower closed set $\widehat{\Omega}$ to define the function $B:\Xx\to \real$ as follows:
\begin{align}\label{eq:B}
B(x) = \begin{cases}\gamma & x\in \widehat{\Omega}\\
\eta+1 & x\not\in \widehat{\Omega}\end{cases}    
\end{align}
The next lemma proves a useful property of the function $B$. 
\begin{lemma}\label{lem:Bmon}
Consider a monotone dtDS $\Sys=(\Xx,\Xx_0,U,f)$ with the lower closed set $\widehat{\Omega}$ as defined in~\eqref{eq:reachset-lower}. Then the function $B:\Xx\to \real$ defined in~\eqref{eq:B} is monotone.
\end{lemma}
\begin{proof}
    Let $x,y\in \Xx$ be such that $x\le y$. We consider two cases i) $y\in \widehat{\Omega}$ and ii) $y\not\in \widehat{\Omega}$. 
    \begin{enumerate}
        \item[i)] In this case, by definition of $\widehat{\Omega}$ in~\eqref{eq:reachset-lower}, there exists a point $z\in \Omega$ such that $y\le z$. Moreover, we have $x\le y\le z$. Therefore, we have $x\in \widehat{\Omega}$ and $B(x) = \gamma = B(y)$.    
        \item[ii)] In this case, we have $B(y) = \eta+1$. Since we know that $\gamma\le \eta< \eta+1$, we get $B(x) \le \eta + 1 = B(y)$.  
    \end{enumerate}
Thus, in both cases i) and ii), we have $B(x)\le B(y)$. This implies that $B$ is a monotone map. 
\end{proof}

Now we go back to the proof of Theorem~\ref{thm:monbar} part~\ref{p1}. We divide the proof into \textbf{if} part and \textbf{only if} part.

\paragraph{Proof of \textit{if part}.}
Suppose that $B:\Xx\to\real$ is a monotone
function satisfying conditions~\eqref{eq:indbar1}, \eqref{eq:indbar2}, and \eqref{eq:indbar3}.
Fix any $x_0\in\Xx_0$ and any admissible input sequence $\{u_t\}_{t=0}^{\infty}$, with $x_{t+1}=f(x_t,u_t)$.
By condition~\eqref{eq:indbar1}, $B(x_t)\le\gamma\le\delta$ for every
$t\in[K-1]$. Applying condition~\eqref{eq:indbar3} gives
$B(x_K)\le\delta$. Since the shifted sequence $x_1,\ldots,x_K$ also satisfies the antecedent of~\eqref{eq:indbar3}, we obtain $B(x_{K+1})\le\delta$. Repeating this argument yields $B(x_t)\le\delta$ for all $t\ge0$. On the other hand, by
condition~\eqref{eq:indbar2}, $B(x)\ge\eta>\delta$ for every $x\in\Xx_u$.
Thus, $\{x_t\}_{t=0}^{\infty}\cap\Xx_u
=\emptyset$. Since $x_0$ and the input sequence were arbitrary, $\Sys$ is robustly safe with respect to $\Xx_u$.

\paragraph{Proof of \textit{only if part}.} Now assume that the monotone dtDS $\Sys$ is safe with respect to upper closed unsafe set $\Xx_u$, for every input $\{u_t\}_{t=0}^{\infty}\in U^{\mathbb{N}}$. Let $\widehat{\Omega}$ be the lower closed set defined in~\eqref{eq:reachset-lower} and $B:\Xx\to \real$ be the function defined in~\eqref{eq:B}. 
Our goal is to show that $B:\Xx\to \real$ defined in~\eqref{eq:B} is a monotone $K$-inductive robust barrier certificate for $\Sys$ satisfying conditions~\eqref{eq:indbar1},\eqref{eq:indbar2}, and~\eqref{eq:indbar3}. 
First, by Lemma~\ref{lem:Bmon}, the function $B:\Xx\to \real$ is monotone.
\begin{enumerate}
    \item\textbf{($B$ satisfies condition~\eqref{eq:indbar1})} Suppose that $x\in \Xx_0$. By the definition of $\Omega$, this implies that $f^{k}(x,\mathbf{u}) \in \Omega$, for every $k\in [K-1]$ and every $\mathbf{u}\in U^{K}$. As a result, we have $f^{k}(x,\mathbf{u})\in \Omega \subseteq \widehat{\Omega}$, for every $k\in [K-1]$ and every $\mathbf{u}\in U^{K}$. Using the definition of $B$ in~\eqref{eq:B}, we get $B(f^{k}(x,\mathbf{u})) = \gamma$, for every $k\in [K-1]$ and every $\mathbf{u}\in U^{K}$. This shows that $B$ satisfies condition~\eqref{eq:indbar1}.
    
    \item\textbf{($B$ satisfies condition~\eqref{eq:indbar2})} Suppose that $x\in \Xx_u$. Since $\Sys$ is safe with respect to $\Xx_u$ for any input $\{u_t\}_{t=0}^{\infty}\in U$, we get that $x\not\in \Omega$. Note that $\widehat{\Omega}$ is the smallest lower closed set containing $\Omega$ and $\Xx_u$ is an upper closed set. As a result, $x\not\in \widehat{\Omega}$. By definition of $B$ in~\eqref{eq:B}, we have $B(x) = \eta + 1 \ge \eta$. This shows that $B$ satisfies condition~\eqref{eq:indbar2}.

    \item \textbf{($B$ satisfies condition~\eqref{eq:indbar3})}
    Let $x\in\Xx$ and $\mathbf{u}\in U^{K-1}$ be such that $B(f^{k}(x,\mathbf{u}))\le\delta$, for every $k\in[K-1]$. By the definition of $B$ in~\eqref{eq:B}, this implies that $f^{k}(x,\mathbf{u})\in \widehat{\Omega}$, for every $k\in[K-1]$. In particular, by the definition of $\widehat{\Omega}$
    in~\eqref{eq:reachset-lower}, there exists $y\in\Omega$ such that $f^{K-1}(x,\mathbf{u})\le y$. Since the state-transition map $f$ is monotone,
    for every $u\in U$, we have $f(f^{K-1}(x,\mathbf{u}),u)\le f(y,u)$.
    Moreover, since $y\in\Omega$, Lemma~\ref{lem:reach}
    implies that $f(y,u)\in\Omega$ for every $u\in U$. Thus, by the definition of $\widehat{\Omega}$
    in~\eqref{eq:reachset-lower}, we obtain
    $f(f^{K-1}(x,\mathbf{u}),u)\in\widehat{\Omega}$.
    Using the definition of $B$ in~\eqref{eq:B}, we conclude that
    $B(f(f^{K-1}(x,\mathbf{u}),u))=\gamma\le\delta$,
    for every $u\in U$. Since $x$ and $\mathbf{u}$ were arbitrary, $B$ satisfies condition~\eqref{eq:indbar3}.

\end{enumerate}
This completes the proof of Theorem~\ref{thm:monbar}~part~\ref{p1}.

\section{Proof of Theorem~\ref{thm:datamonbar}}\label{App:proofdata1}

Regarding part~\ref{thm:datamonbar1}, suppose that
$\mathcal{B}:\Xx\to\mathbb{R}$ and $\kappa:\Xx\to U$
are monotone functions satisfying
conditions~\eqref{eq:data1-c}--\eqref{eq:data3-c}.
Define the closed-loop transition map
\[
F_\kappa(x):=f(x,\kappa(x)).
\]
Since $f$ and $\kappa$ are monotone, $F_\kappa$ is
monotone. Under the notation introduced in the paper,
$F_\kappa^k(x)=f^k(x,\kappa(x))$.
Our goal is to establish the $K$-inductive barrier
conditions for this closed-loop system.

\begin{enumerate}
\item \textbf{(Initial-set condition)}
Let $x\in\Xx_0$. By~\eqref{eq:cover}, there exists
$i\in\mathcal{I}$ such that
$x\in[\underline{x}^i,\bar{x}^i]$.
By monotonicity of $\mathcal{B}$ and $F_\kappa$,
and condition~\eqref{eq:data1-c}, we obtain
\[
\mathcal{B}(F_\kappa^k(x))
\le
\mathcal{B}(F_\kappa^k(\bar{x}^i))
\le \gamma,
\qquad \forall k\in[K-1].
\]

\item \textbf{(Unsafe-set condition)}
Let $x\in\Xx_u$. By~\eqref{eq:cover}, there exists
$j\in\mathcal{U}$ such that
$\underline{x}^j\le x$.
Condition~\eqref{eq:data2-c} and monotonicity of
$\mathcal{B}$ imply
\[
\eta
\le \mathcal{B}(\underline{x}^j)
\le \mathcal{B}(x).
\]

\item \textbf{(Closed-loop induction condition)}
Let $x\in\Xx$ satisfy
\[
\max_{k\in[K-1]}
\mathcal{B}(F_\kappa^k(x))\le\delta.
\]
Choose $i\in\mathcal{P}$ such that
$x\in[\underline{x}^i,\bar{x}^i]$.
By monotonicity of $\mathcal{B}$ and $F_\kappa$,
\[
\max_{k\in[K-1]}
\mathcal{B}(F_\kappa^k(\underline{x}^i))
\le
\max_{k\in[K-1]}
\mathcal{B}(F_\kappa^k(x))
\le\delta.
\]
Condition~\eqref{eq:data3-c} therefore gives
$\mathcal{B}(F_\kappa^K(\bar{x}^i))\le\delta$.
Using monotonicity once more, we obtain
\[
\mathcal{B}(F_\kappa^K(x))
\le
\mathcal{B}(F_\kappa^K(\bar{x}^i))
\le\delta.
\]
\end{enumerate}

Now fix any $x_0\in\Xx_0$ and consider the closed-loop
trajectory $x_{t+1}=F_\kappa(x_t)$.
The initial-set condition gives
$\mathcal{B}(x_t)\le\gamma\le\delta$
for every $t\in[K-1]$.
The closed-loop induction condition then yields
$\mathcal{B}(x_K)\le\delta$.
Repeated application of this condition to successive
windows of $K$ states gives
$\mathcal{B}(x_t)\le\delta$ for all $t\ge0$.
Since $\mathcal{B}(x)\ge\eta>\delta$ for every
$x\in\Xx_u$, the trajectory never enters $\Xx_u$.
Thus, $\Sys$ is safely controllable with respect to
$\Xx_u$ under the feedback controller $\kappa$.

Regarding part~\ref{thm:datamonbar2}, suppose that $\mathcal{B}$ is a monotone function satisfying conditions~\eqref{eq:data1}-\eqref{eq:data3}. Our goal is to show that $\mathcal{B}$ is a monotone robust $K$-inductive barrier certificate for $\Sys$, i.e., it satisfies all the conditions~\eqref{eq:indbar1}-\eqref{eq:indbar3}.

\begin{enumerate}
\item \textbf{($\mathcal{B}$ satisfies condition~\eqref{eq:indbar1})} Let $x\in\Xx_0$. Since
$\Xx_0\subseteq \bigcup_{i\in\mathcal I}[\underline{x}^i,\bar{x}^i]$,
there exists $i\in\mathcal I$ such that
$x\le \bar{x}^i$.
Let $\mathbf u\in U^K$. Since
$U\subseteq \bigcup_{j\in\mathcal Q}[\underline{u}^j,\bar{u}^j]$,
there exists
$\bar{\mathbf u}\in
\left(
\bigcup_{j\in\mathcal Q}\{\bar{u}^j\}
\right)^K$
such that
$\mathbf u\le \bar{\mathbf u}$.
Since $\mathcal{B}$ satisfies condition~\eqref{eq:data1}, we have
\[
\mathcal{B}(f^k(\bar{x}^i,\bar{\mathbf u}))\le \gamma,
\qquad \forall k\in[K-1].
\]
On the other hand, $\mathcal{B}$ is monotone and $f$ is a monotone state-transition map. Thus,
\[
\mathcal{B}(f^k(x,\mathbf u))
\le
\mathcal{B}(f^k(\bar{x}^i,\mathbf u))
\le
\mathcal{B}(f^k(\bar{x}^i,\bar{\mathbf u}))
\le \gamma,
\qquad \forall k\in[K-1].
\]
This shows that $\mathcal{B}$ satisfies condition~\eqref{eq:indbar1}.

\item \textbf{($\mathcal{B}$ satisfies condition~\eqref{eq:indbar2})} Let $x\in\Xx_u$. Since
$\Xx_u\subseteq \bigcup_{i\in\mathcal U}[\underline{x}^i,\bar{x}^i]$,
there exists $j\in\mathcal U$ such that
$\underline{x}^j\le x$.
Since $\mathcal{B}$ satisfies condition~\eqref{eq:data2}, we have
$\eta\le \mathcal{B}(\underline{x}^j)$.
Using monotonicity of $\mathcal{B}$, we get
\[
\eta\le \mathcal{B}(\underline{x}^j)\le \mathcal{B}(x).
\]
Hence, $\mathcal{B}$ satisfies condition~\eqref{eq:indbar2}.

\item \textbf{($\mathcal{B}$ satisfies condition~\eqref{eq:indbar3})} Let $x\in\Xx$ and $\mathbf u\in U^{K+1}$ be such that
$\mathcal{B}(f^k(x,\mathbf u))\le \delta,
\;\; \forall k\in[K-1]$.
Since the family of hyper-rectangles
$\{[\underline{x}^i,\bar{x}^i]\}_{i\in\mathcal P}$
covers $\Xx$, there exists $i\in\mathcal P$ such that
$x\in[\underline{x}^i,\bar{x}^i]$.
Also, there exist
\[
\bar{\mathbf u}\in
\left(
\bigcup_{j\in\mathcal Q}\{\bar{u}^j\}
\right)^{K+1},
\qquad
\underline{\mathbf u}\in
\left(
\bigcup_{j\in\mathcal Q}\{\underline{u}^j\}
\right)^{K+1},
\]
such that
$\underline{\mathbf u}\le \mathbf u\le \bar{\mathbf u}$.
Using monotonicity of $\mathcal{B}$ and $f$, we get
\[
\mathcal{B}(f^k(\underline{x}^i,\underline{\mathbf u}))
\le
\mathcal{B}(f^k(x,\mathbf u))
\le \delta,
\qquad \forall k\in[K-1].
\]
Since $\mathcal{B}$ satisfies condition~\eqref{eq:data3}, we have
$\mathcal{B}(f^K(\bar{x}^i,\bar{\mathbf u}))\le \delta$.
As a result,
\[
\mathcal{B}(f^K(x,\mathbf u))
\le
\mathcal{B}(f^K(\bar{x}^i,\mathbf u))
\le
\mathcal{B}(f^K(\bar{x}^i,\bar{\mathbf u}))
\le \delta,
\]
where the first inequality holds by monotonicity of $N$ and $f$ together with $x\le \bar{x}^i$, and the second inequality holds by monotonicity of $\mathcal{B}$ and $f$ together with $\mathbf u\le \bar{\mathbf u}$. Hence, $\mathcal{B}$ satisfies condition~\eqref{eq:indbar3}.
\end{enumerate}
Therefore, $\mathcal{B}$ satisfies conditions~\eqref{eq:indbar1}--\eqref{eq:indbar3}. Hence, $\mathcal{B}$ is a $K$-inductive robust barrier certificate. The result then follows from Theorem~\ref{thm:monbar} part~\ref{p1}. Therefore, the system $\mathfrak S$ is robustly safe with respect to $\Xx_u$. This completes the proof.


\section{Memory-efficient inductive barrier certificates}\label{app:me}

In this appendix, by leveraging the monotonicity of the barrier certificate, we propose the following memory-efficient formulation of control barrier certificates. Let $\gamma,\delta,\eta\in\mathbb{R}$ satisfy $\gamma\leq\delta<\eta$. We have:
\begin{align}
&\mathcal{B}(f(\underline{x}^i,\kappa(\underline{x}^i)) \geq \mathcal{B}(\bar{x}^i)
&&\forall i \in \mathcal{P},\label{eq:data1-cw} \\
&\mathcal{B}(f^{K-1}(\bar{x}^i,\kappa(\bar{x}^i))) \leq \gamma && \forall i \in \mathcal{I},\\
&\mathcal{B}(\underline{x}^i) \ge \eta  &&\forall i \in \mathcal{U},\label{eq:data2-cw}\\
&\mathcal{B}(f^{K-1}(\underline{x}^i,\kappa(\underline{x}^i))) \leq \delta \; \Rightarrow \; \mathcal{B}(f^{K}(\bar{x}^i,\kappa(\bar{x}^i)))\leq \delta, &&\forall i\in \mathcal{P}\label{eq:data3-cw}   
\end{align}
Similarly, we can propose the following memory-efficient formulation of robust barrier certificates:
\begin{align}
&\mathcal{B}(f(\underline{x}^i,\underline{u}^j)) \geq \mathcal{B}(\bar{x}^i)  &&\forall i \in \mathcal{P},\;\forall j \in \mathcal{Q}, \label{eq:data1-w}\\
&\mathcal{B}(f^{K-1}(\bar{x}^i,\overline{\mathbf{u}})) \leq \gamma &&\forall i \in \mathcal{I}, \; \forall \overline{\mathbf{u}}\in \overline{\mathcal{U}}^{K-1}\\
&\mathcal{B}(\underline{x}^i) \ge \eta  &&\forall i \in \mathcal{U},\label{eq:data2-w}\\
&\mathcal{B}(f^{K-1}(\underline{x}^i,\underline{\mathbf{u}})) \leq \delta \; \Rightarrow \; \mathcal{B}(f^{K}(\bar{x}^i,\overline{\mathbf{u}}))\leq \delta, && \forall i\in \mathcal{P},\; \forall \underline{\mathbf{u}}\in \underline{\mathcal{U}}^{K-1},\; \forall \overline{\mathbf{u}}\in \overline{\mathcal{U}}^{K}\label{eq:data3-w}   
\end{align}
where $\overline{\mathcal{U}} = 
\bigcup_{j\in\mathcal Q}\{\bar{u}^j\}$ and $\underline{\mathcal{U}} = 
\bigcup_{j\in\mathcal Q}\{\underline{u}^j\}$. The following results demonstrate the effectiveness of the memory-efficient formulation of control and robust barrier certificates.

\begin{corollary}[Memory-efficient inductive barrier certificates]\label{cor}
    Consider a monotone dtDS $\mathfrak{S} = (\mathcal{X}, \mathcal{X}_0, U, f)$ with an unknown state-transition map $f$ and a set of unsafe states $\mathcal{X}_u$. Let $\{[\underline{x}^i, \bar{x}^i]\}_{i \in \mathcal{P}}$ be a family of hyper-rectangles covering the state space $\mathcal{X}$, and let $\{[\underline{u}^j, \bar{u}^j]\}_{j \in \mathcal{Q}}$ be a family of hyper-rectangles covering the input space $U$. Define the index sets $\mathcal{I}$ and $\mathcal{U}$ as the minimal subsets of $\mathcal{P}$ satisfying \eqref{eq:cover}. Then the system $\mathfrak{S}$

\begin{enumerate}
\item \emph{(Safe Control Synthesis)} is safely controllable with respect to $\mathcal{X}_u$ if there exist monotone inductive barrier certificate $\mathcal{B}: \mathcal{X} \to \real$ and a monotone feedback $\kappa : \mathcal{X} \to U$ satisfying conditions~\eqref{eq:data1-cw}--\eqref{eq:data3-cw}.
\item\label{p1:me} \emph{(Robust Safety Verification)} is robustly safe with respect to $\mathcal{X}_u$ if there exists a monotone inductive barrier certificate $\mathcal{B}:\mathcal{X}\to \mathbb{R}$ satisfying conditions~\eqref{eq:data1-w}--\eqref{eq:data3-w}.
\end{enumerate}
\end{corollary}

Corollary~\ref{cor} enables safe controller synthesis and verification of safety with reduced memory requirements and improved computational efficiency. Therefore, throughout this paper, we can replace \ref{eq:data1}--\ref{eq:data3} with \ref{eq:data1-w}--\ref{eq:data3-w}, or \ref{eq:data1-c}--\ref{eq:data3-c} with \ref{eq:data1-cw}--\ref{eq:data3-cw}, to obtain a memory-efficient version of the robust barrier or safe controller.

\section{Inductive vs. classic monotone barriers}\label{App:compbarr}
We consider a discrete-time freeway traffic system~\cite{FB-CTDS} $\mathfrak{S}=(\mathcal{X},\mathcal{X}_0,f)$ with state $q(t)=[q_1(t),q_2(t)]^\top$, where $q_i(t)$ denotes the vehicle density in road segment $i$ at time $t$. The state space is $\mathcal{X}=[0,50]\times[0,50]$. We fix the sampling time to $\tau=0.5$ and the upstream inflow to $F_{0\to 1}=8.3$. The flow from segment $1$ to segment $2$ is given by the demand--supply rule $F_{1\to 2}(q(t))=\min\{\varphi(q_1(t)),\,\sigma(q_2(t))\}$, where the demand and supply functions are defined as $\varphi(x)=10\big(1-e^{-0.1x}\big)$ and $\sigma(x)=\max\{0,\,10\big(1-(x/50)^3\big)\}$. The resulting discrete-time dynamics are
\[
q_1(t+1)=q_1(t)+\tau\big(F_{0\to 1}-F_{1\to 2}(q(t))\big),\qquad
q_2(t+1)=q_2(t)+\tau\big(F_{1\to 2}(q(t))-\varphi(q_2(t))\big),
\]
followed by component-wise saturation to enforce $q(t+1)\in\mathcal{X}$. The initial set is $\mathcal{X}_0=[0,10]\times[0,10]$, and the unsafe set is defined as $\mathcal{X}_u=\mathcal{X}\setminus\big([0,40]\times[0,40]\big)$, which captures congestion whenever at least one segment density exceeds $40$. 

\begin{figure}[!h]
 \centering
  \includegraphics[width =0.6\linewidth]{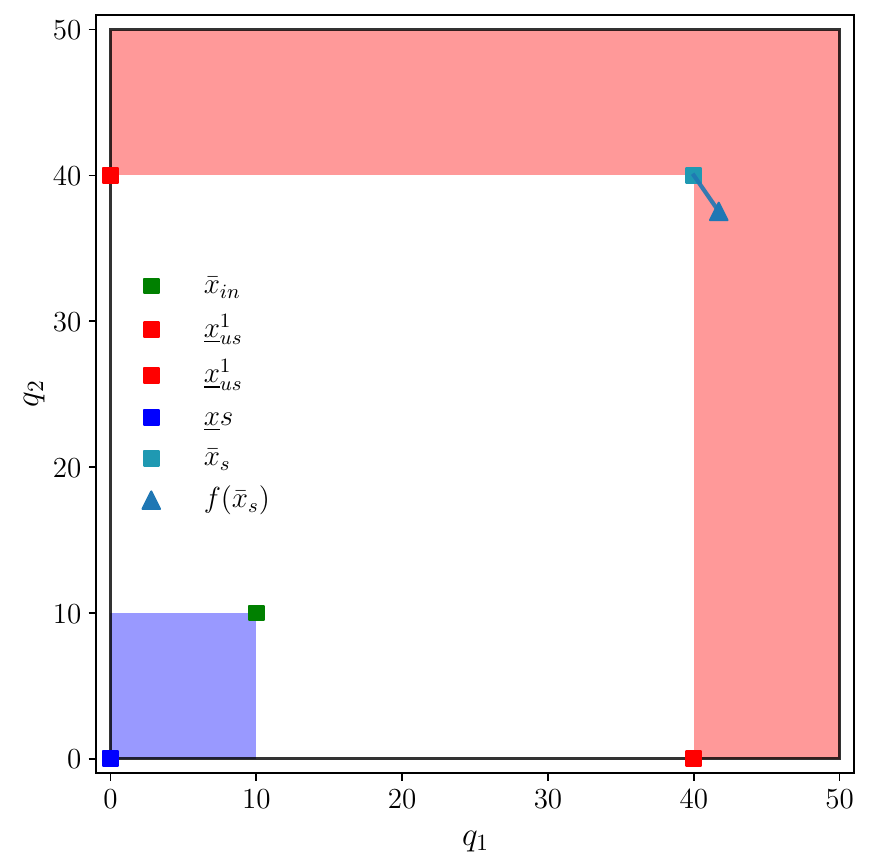}
  \caption{Extreme-point evaluation of one-step dynamics.}
	\label{fig:tr2df}
 \end{figure}

To verify safety using only sampled data, we employ a fixed hyper-rectangular gridding of the state space. Specifically, we partition the unsafe region into two rectangles, namely $[0,40]\times[40,50]$ and $[40,50]\times[0,50]$, and define one rectangle for the initial set $[0,10]\times[0,10]$, together with a larger rectangle $[0,40]\times[0,40]$ covering the remaining safe domain. This results in a total of four hyper-rectangles and five representative corner points $\{ \overline{x}_{in},\underline{x}^1_{us},\underline{x}^2_{us},\underline{x}_{s},\overline{x}_{s} \}$. For the corner point $\overline{x}_{s}$, we generate trajectory samples by evaluating the one-step transition $f(\overline{x}_{s})$. Figure~\ref{fig:tr2df} illustrates the resulting rectangular gridding, where the sampled points $\{ \overline{x}_{in},\underline{x}^1_{us},\underline{x}^2_{us},\underline{x}_{s},\overline{x}_{s} \}$ are marked with squares and $f(\overline{x}_{s})$ is marked with triangle.

\begin{figure}[!h]
 \centering
  \includegraphics[width =0.6\linewidth]{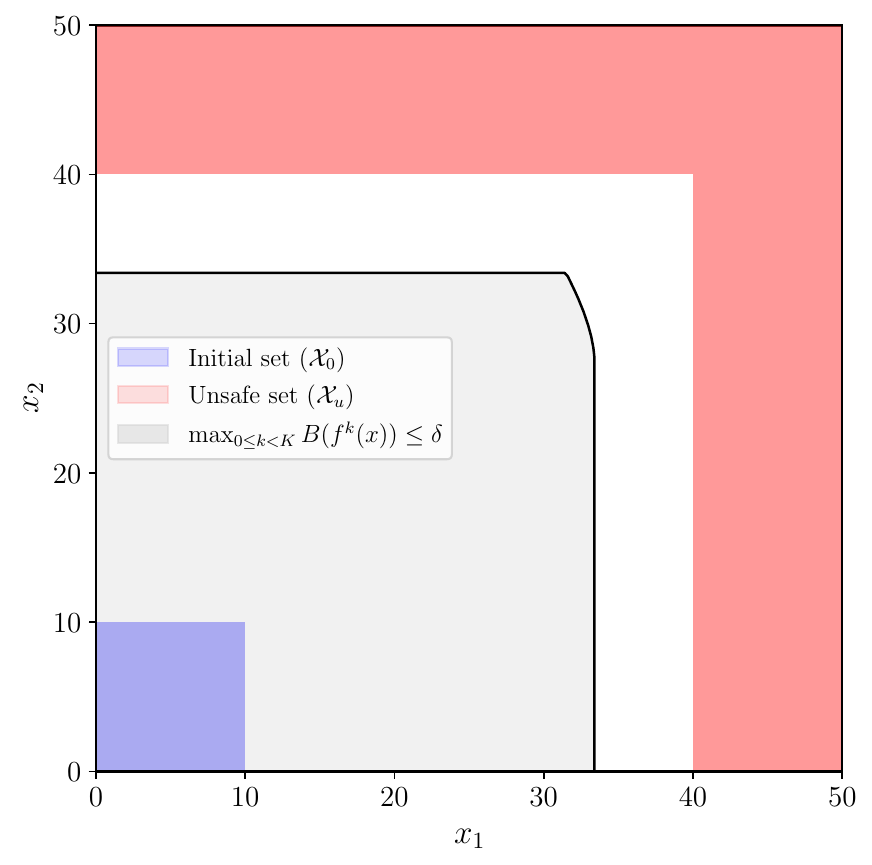}
  \caption{Sub-level set of the learned neural inductive barrier certificate over the state space. The parameters $\gamma = 5$, $\delta = 9.5$, and $\eta = 10$ are used in this experiment.}
	\label{fig:indtraf}
 \end{figure}

First, we train a neural inductive barrier certificate using the sampled data
collected from this grid and verify safety via Algorithm~\ref{alg:learningrobustwithshrinking}. Using a simplified Max-Monotone Network based on the direct max-input branch in Definition~\ref{eq:monnn}, with a monotone hidden layer of 5 neurons and $K=100$, our method successfully verifies the safety of the 2-D traffic system in $0.05\,\mathrm{s}$. Figure~\ref{fig:indtraf} illustrates the resulting sub-level set of the learned neural inductive barrier over the state space. In contrast, a classic ($K=1$) monotone neural barrier certificate cannot certify safety under the same fixed gridding. In particular, although $B(\underline{x}_s) \leq 0$ holds since $\underline{x}_s \in \Xx_0$, we have $B\!\left(f(\overline{x}_s)\right) > 0$, which violates the barrier condition. Consequently, safety cannot be verified using an ordinary monotone barrier on this coarse partition.

\section{Max-Monotone Network}

In this appendix, we present additional theoretical results and technical details concerning the Max-Monotone Networks introduced in Definition~\ref{def:monNN}.

\subsection{Max-separable inductive barrier certificates}\label{App:justminmax}

In this appendix, we show that, under mild assumptions, the existence of a monotone inductive barrier certificate for a monotone system $\Sys=(\Xx,\Xx_0,U,f)$ implies the existence of a min-max-separable inductive barrier certificate of the form~\eqref{eq:min-max-separable} where  $h_{i,j}:\real\to\real$ is monotone.

\begin{theorem}[Min--max-separable monotone
$K$-inductive barrier certificates]
\label{thm:min_max_separable}
Consider a monotone dynamical system
$\mathcal{S}=(\mathcal{X},\mathcal{X}_0,U,f)$ with $\mathcal{X}\subseteq \real^n$ and with an
unsafe set $\mathcal{X}_u\subseteq\mathcal{X}$. For $K\geq 1$, suppose that there exists a monotone controlled
$K$-inductive barrier certificate
$B:\mathcal{X}\to\mathbb{R}$ satisfying conditions~\eqref{eq:indbar1-c}--\eqref{eq:indbar3-c}. Then there exists a monotone controlled
$K$-inductive barrier certificate
$\widetilde{B}:\mathcal{X}\to\mathbb{R}$ for system $\mathfrak{S}$ satisfying conditions~\eqref{eq:indbar1-c}--\eqref{eq:indbar3-c}
that can be expressed as the pointwise minimum
of a family of max-separable\footnote{A function $f:\real^n\to \real$ is max-separable, if there exist scalar functions $f_i:\real\to \real$ such that $f(x) = \max_{i\in \{1,\ldots,n\}}f_i(x_i)$.} monotone functions:
\begin{equation}
\label{eq:min_max_barrier}
\widetilde{B}(x)
=
\min_{j\in J}\left\{
\max_{i\in \{1,\ldots,n\}} h_{i,j}(x_i)\right\},
\end{equation}
where $h_{i,j}:\mathbb{R}\to\mathbb{R}$
is a nondecreasing function and $J$ is an index set.
\end{theorem}

\begin{proof}
Define the $\delta$-sublevel set of $B$ by
\begin{equation}
\label{eq:C_delta}
\mathcal{C}_{\delta}
:=
\{x\in\mathcal{X}:B(x)\leq\delta\}.
\end{equation}

We first consider the case
$\mathcal{C}_{\delta}=\varnothing$.
In this case, condition~\eqref{eq:indbar1-c} implies
$\mathcal{X}_0=\varnothing$, and the antecedent
of the condition~\eqref{eq:indbar3-c} is never satisfied.
Consequently, the constant function
$\widetilde{B}(x)=\eta+1$
satisfies the barrier conditions~\eqref{eq:indbar1-c}--\eqref{eq:indbar3-c}.
It also admits the representation
\eqref{eq:min_max_barrier} with a singleton
index set $J$ and constant nondecreasing function. Now, assume that
$\mathcal{C}_{\delta}\neq\varnothing$. Since $B$ is monotone, $\mathcal{C}_{\delta}$
is lower closed relative to $\mathcal{X}$.
In particular, for any $x,y\in\mathcal{X}$,
\begin{equation}
x\leq y,\quad y\in\mathcal{C}_{\delta}
\quad\Longrightarrow\quad
x\in\mathcal{C}_{\delta}.
\label{eq:lower_closed}
\end{equation}
For each $y\in\mathcal{C}_{\delta}$ and
$i\in \{1,\ldots,n\}$, define $h_{i,y}:\real\to \real$ by
\begin{equation}
\label{eq:scalar_step}
h_{i,y}(s)
:=
\begin{cases}
\gamma, & s\leq y_i,\\
\eta+1, & s>y_i.
\end{cases}
\end{equation}
Each function $h_{i,y}$ is nondecreasing
because $\gamma\leq\delta<\eta+1$.
Now define
\begin{equation}
\label{eq:max_separable_component}
G_y(x)
:=
\max_{i\in \{1,\ldots,n\}}h_{i,y}(x_i).
\end{equation}

By construction,
\begin{equation}
\label{eq:G_y_characterization}
G_y(x)
=
\begin{cases}
\gamma, & x\leq y,\\
\eta+1, & x\not\leq y.
\end{cases}
\end{equation}

Define
\begin{equation}
\label{eq:corrected_barrier}
\widetilde{B}(x)
:=
\min_{y\in\mathcal{C}_{\delta}}G_y(x)
=
\min_{y\in\mathcal{C}_{\delta}}\left\{
\max_{i\in \{1,\ldots,n\}}h_{i,y}(x_i)\right\}.
\end{equation}
The minimum is well defined because every
$G_y$ takes values in the two-element set
$\{\gamma,\eta+1\}$.
We show that
\begin{equation}
\label{eq:barrier_characterization}
\widetilde{B}(x)
=
\begin{cases}
\gamma,
& x\in\mathcal{C}_{\delta},\\
\eta+1,
& x\notin\mathcal{C}_{\delta}.
\end{cases}
\end{equation}

Indeed, if $x\in\mathcal{C}_{\delta}$,
we may choose $y=x$, which gives
$G_y(x)=\gamma$.
Since $G_y(x)\geq\gamma$ for every $y$,
we conclude that
$\widetilde{B}(x)=\gamma$.
Conversely, suppose
$x\notin\mathcal{C}_{\delta}$.
For every $y\in\mathcal{C}_{\delta}$,
we must have $x\not\leq y$.
Otherwise, the lower-closedness of
$\mathcal{C}_{\delta}$ would imply
$x\in\mathcal{C}_{\delta}$,
a contradiction.
Thus, $G_y(x)=\eta+1$ for every
$y\in\mathcal{C}_{\delta}$, and hence
$\widetilde{B}(x)=\eta+1$.
This establishes
\eqref{eq:barrier_characterization}.

We next verify that $\widetilde{B}$ satisfies the barrier conditions~\eqref{eq:indbar1-c}--\eqref{eq:indbar3-c}.

\begin{enumerate}
\item \textbf{($\widetilde{B}$ satisfies condition~\eqref{eq:indbar1-c})} Since $B$ is a barrier certificate
for every $x\in\mathcal{X}_0$, there exists $\mathbf{u}\in U^{K-1}$ such that $\max_{k\in [K-1]}B(f^k(x,\mathbf{u}))\leq\gamma\leq\delta$. This implies that, for every $k\in [K-1]$ and $x\in \mathcal{X}_0$, there exists $\mathbf{u}\in U^{K-1}$ such that $f^k(x,\mathbf{u}) \in C_{\delta}$.
Therefore, by \eqref{eq:barrier_characterization}, for every $x\in\mathcal{X}_0$, there exists $\mathbf{u}\in U^{K-1}$ such that $\max_{k\in [K-1]}\widetilde{B}(f^k(x,\mathbf{u}))=\gamma\le \gamma$.

\item \textbf{($\widetilde{B}$ satisfies condition~\eqref{eq:indbar2-c})} Since $B$ satisfies the unsafe-set
condition and $\eta>\delta$, we have
$\mathcal{X}_u\cap\mathcal{C}_{\delta}
=\varnothing$.
It follows from
\eqref{eq:barrier_characterization} that $\widetilde{B}(x)=\eta+1>\eta$, for every $x\in\mathcal{X}_u$.

\item \textbf{($\widetilde{B}$ satisfies condition~\eqref{eq:indbar3-c})} Using the characterization of $\widetilde{B}$ in~\eqref{eq:barrier_characterization}, one can easily check that
\begin{equation}
\label{eq:sublevel_equivalence}
\widetilde{B}(x)\leq\delta
\quad\Longleftrightarrow\quad
x\in\mathcal{C}_{\delta}
\quad\Longleftrightarrow\quad
B(x)\leq\delta.
\end{equation}

This means that, the $\delta$-sublevel sets
of $B$ and $\widetilde{B}$ coincide.
Consequently, any admissible control choices that
establish the condition~\eqref{eq:indbar3-c} for $B$ also establish
the condition~\eqref{eq:indbar3-c} for
$\widetilde{B}$.

\end{enumerate}

Finally, $h_{i,y}$ is non-decreasing, for every $y\in \mathcal{C}_{\delta}$. This implies that $G_y$ is monotone, for every $y\in \mathcal{C}_{\delta}$. Therefore, their pointwise
minimum $\widetilde{B}$ is also monotone. Thus, $\widetilde{B}$ is a monotone controlled $K$-inductive barrier certificate for the system $\mathfrak{S}$
admitting the representation
\eqref{eq:min_max_barrier} for $J=C_{\delta}$.
\end{proof}

\subsection{Architecture of Max-Monotone Networks}\label{App:proofminmax}
\label{App:NNarch}

\begin{figure}[t]
\centering
\includegraphics[width=0.98\linewidth]{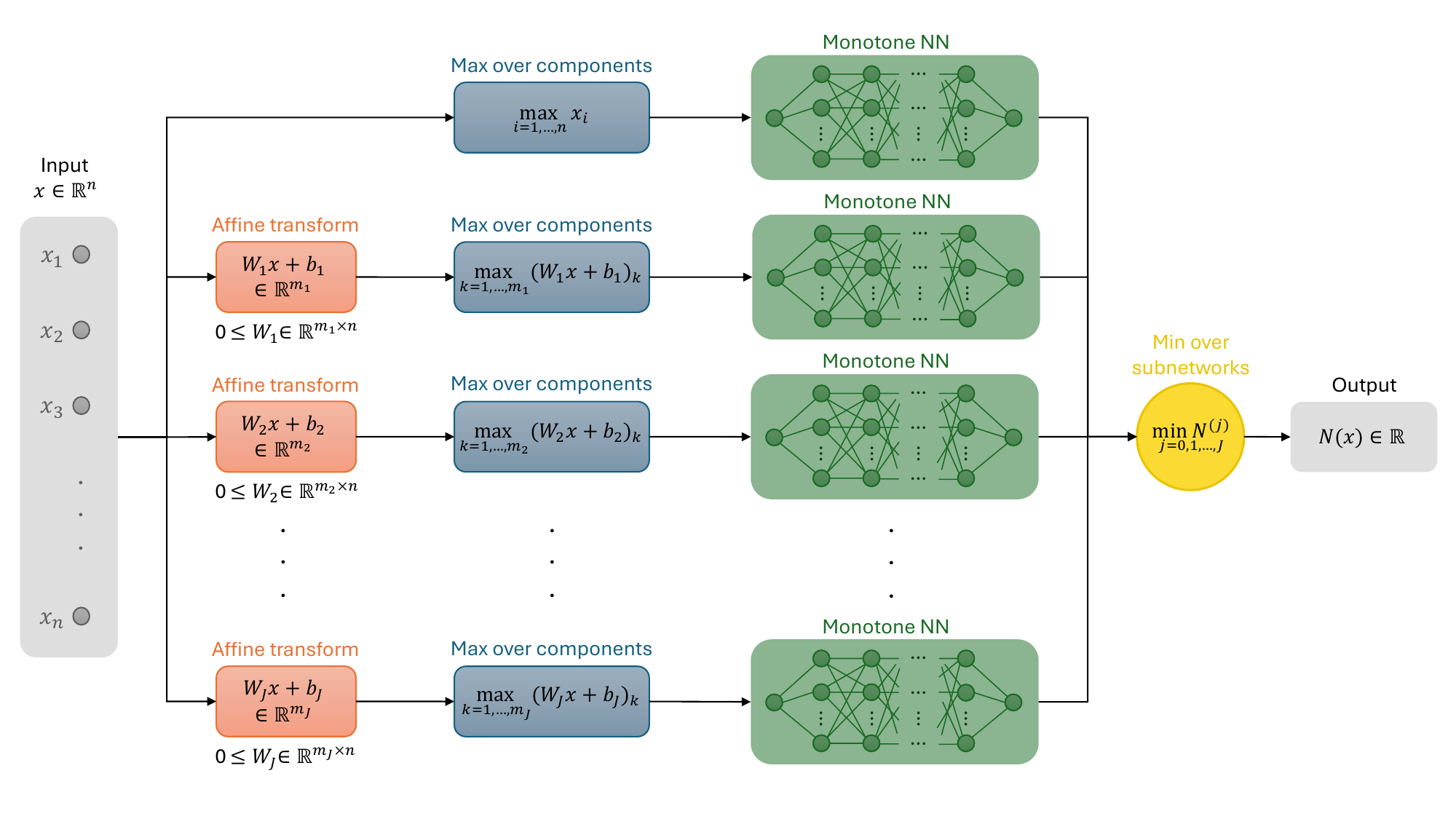}
\caption{Schematic illustration of the proposed Max-Monotone Network architecture.}
\label{fig:minmax_arch}
\end{figure}

Figure~\ref{fig:minmax_arch} provides a schematic illustration of the proposed Max-Monotone Network architecture defined in Section~\ref{subsec:monNN}. The network consists of a direct max-input branch and multiple affine max branches, each followed by a monotone sub-network, with the final output obtained via a minimum operator.

\subsection{Proof of Theorem~\ref{thm:minmaxnn}}\label{app:minmaxnnproof}
We first establish monotonicity, and then prove the universal approximation property.

\emph{Monotonicity part.}
Let $x,y \in \mathbb{R}^n$ such that $x \le y$ componentwise. We show that $N(x) \le N(y)$.

First, for $j=0$, we have
\[
g^{(0)}(x) = \max_i x_i \le \max_i y_i = g^{(0)}(y).
\]
For $j=1,\ldots,J$, since $W_j \ge 0$, it follows that $W_j x \le W_j y$ componentwise, and hence
\[
(W_j x + b_j)_k \le (W_j y + b_j)_k, \quad \forall k.
\]
Taking the maximum over $k$ yields
\[
g^{(j)}(x) \le g^{(j)}(y), \quad \forall j=1,\ldots,J.
\]
Since each $N^{(j)}$ is a monotone function, we obtain
\[
N^{(j)}\big(g^{(j)}(x)\big) \le N^{(j)}\big(g^{(j)}(y)\big), \quad \forall j=0,\ldots,J.
\]
Taking the minimum over $j$ preserves the inequality, and therefore
\[
N(x) = \min_j N^{(j)}(g^{(j)}(x)) \le \min_j N^{(j)}(g^{(j)}(y)) = N(y).
\]
Thus, $N$ is monotone.

\emph{Universal approximation part.}
Let $m:[0,1]^n \to \mathbb{R}$ be continuous, bounded, and monotone with bounded partial derivatives. Fix $\varepsilon>0$.
Let
$a = \min_{x\in[0,1]^n} m(x), \; b = \max_{x\in[0,1]^n} m(x)$,
and let $\alpha>0$ be a bound on the magnitude of all partial derivatives of $m$.
We first note that the direct-max branch $N^{(0)}(g^{(0)}(x))$ does not affect the approximation argument. Indeed, choose $N^{(0)}$ to be a constant function satisfying
\[
N^{(0)}(g^{(0)}(x)) > b, \qquad \forall x\in[0,1]^n.
\]
Hence, this branch is never selected by the final minimum. It therefore suffices to construct the approximation using only the max-affine branches $j=1,\ldots,J$.
Partition $[0,1]^n$ into a uniform grid with spacing $\delta = \frac{1}{N}$ along each dimension, and denote the set of grid points by
\[
S = \{(i_1\delta,\ldots,i_n\delta) : 0 \le i_d \le N\}.
\]

For each grid point $x' \in S$, we construct one branch indexed by $j$ and define
\[
g^{(j)}(x) = \max_{k=1,\ldots,n+1} \big((W_jx+b_j)_k\big),
\]
where the affine components are chosen as follows: one component is the constant $m(x')$,
for each coordinate $d=1,\ldots,n$, one component is
\[
m(x') + \gamma(x_d - x'_d),
\]
with $\gamma > \frac{b-a}{\delta}$.
This construction is of the form $g^{(j)}(x)=\max_k((W_jx+b_j)_k)$ with $W_j\ge 0$.
For each such branch, choose the scalar monotone neural network $N^{(j)}$ to be the identity map, i.e.,
\[
N^{(j)}(t)=t.
\]
This is realized exactly by a monotone neural network with a single linear layer of weight $1$ and bias $0$. Hence, for all $j\ge 1$,
\[
N^{(j)}(g^{(j)}(x)) = g^{(j)}(x).
\]

We now show that this construction approximates $m$. Fix $x\in[0,1]^n$, and let $s^{(1)}\in S$ be the largest grid point such that $s^{(1)}\le x$ componentwise, i.e.,
\[
0 \le x_d - s_d^{(1)} < \delta, \qquad \forall d.
\]
Define $s^{(2)}\in S$ componentwise by
$s_d^{(2)}=\min\{s_d^{(1)}+\delta,1\},
\; d=1,\ldots,n$.
Consider the branch associated with $s^{(2)}$. Since $x_d - s_d^{(2)} \le 0$ for all $d$, every affine component in that branch satisfies
\[
m(s^{(2)}) + \gamma(x_d - s_d^{(2)}) \le m(s^{(2)}).
\]
Hence, the maximum is attained at the constant component, and therefore
\[
g^{(j)}(x)=m(s^{(2)}).
\]
Since $N^{(j)}$ is the identity and $N(x)=\min_j N^{(j)}(g^{(j)}(x))$, we obtain
\[
N(x)\le N^{(j)}(g^{(j)}(x)) = g^{(j)}(x)=m(s^{(2)}).
\]

Next, consider any branch associated with a grid point $s' \in S$. If $s'$ dominates $s^{(1)}$, then by monotonicity of $m$,
\[
m(s') \ge m(s^{(1)}).
\]
Since $m(s')$ is one of the affine components in that branch, it follows that
\[
g^{(j)}(x)\ge m(s^{(1)}).
\]
On the other hand, if $s'$ does not dominate $s^{(1)}$, then there exists a coordinate $d$ such that
\[
s'_d < s_d^{(1)}.
\]
Since $s'_d$ and $s_d^{(1)}$ are grid points, we have $s_d^{(1)}-s'_d\ge\delta$.
Together with $x_d\ge s_d^{(1)}$,
this implies 
\[
x_d-s'_d\ge\delta.
\]
Hence, the corresponding affine component satisfies
\[
m(s') + \gamma(x_d - s'_d) \ge a + \gamma\delta > a + (b-a) = b \ge m(s^{(1)}).
\]
Therefore, every branch output is at least $m(s^{(1)})$, and thus
\[
N(x)\ge m(s^{(1)}).
\]

Combining the two inequalities gives
\[
m(s^{(1)}) \le N(x) \le m(s^{(2)}).
\]
Since $m$ is monotone and $s^{(1)} \le x \le s^{(2)}$, we also have
\[
m(s^{(1)}) \le m(x) \le m(s^{(2)}).
\]
Therefore,
\[
|m(x)-N(x)| \le |m(s^{(2)}) - m(s^{(1)})|.
\]

By applying the mean value theorem coordinatewise and using the bounded partial derivatives of $m$, we obtain
\[
|m(s^{(2)})-m(s^{(1)})|
\le \alpha\sum_{d=1}^n
(s_d^{(2)}-s_d^{(1)})
\le n\delta\alpha.
\]
Choosing $\delta < \frac{\varepsilon}{n\alpha}$ yields
\[
|m(x)-N(x)|<\varepsilon, \qquad \forall x\in[0,1]^n.
\]
This completes the proof.

\section{Proof of Theorem~\ref{thm:zeroloss}}\label{App:proofdata}

Regarding part~\ref{thm:zeroloss1}, our goal is to show that if there exist monotone neural networks
$N:\Xx\to\real$ and $\kappa:\Xx\to U$ such that
$L_{\mathrm{total}}^{\mathrm s}
= L_1^{\mathrm s}+L_2^{\mathrm s}+L_3^{\mathrm s}=0$,
then the system $\mathfrak S$ is safely controllable with respect to $\Xx_u$. Since each loss term is nonnegative, we obtain
$L_1^{\mathrm s}=L_2^{\mathrm s}=L_3^{\mathrm s}=0$.
We now show that $(N,\kappa)$ satisfies
conditions~\eqref{eq:data1-c}--\eqref{eq:data3-c}.

\begin{enumerate}
    \item \textbf{($(N,\kappa)$ satisfies condition~\eqref{eq:data1-c})} Since $L_1^{\mathrm s}=0$, every summand in~\eqref{eq:loss1_c} is zero. Hence, for every $i\in\mathcal I$,
\[
\max_{k\in[K-1]} N(f^k(\bar{x}^i,\kappa(\bar{x}^i))) \le \gamma,
\]
which gives~\eqref{eq:data1-c}.

    \item \textbf{($N$ satisfies condition~\eqref{eq:data2-c})}
    Since $L_2^{\mathrm s}=0$, every summand in~\eqref{eq:loss2_c} is zero. Hence, for every $i\in\mathcal U$,
\[
N(\underline{x}^i)\ge \eta,
\]
which gives~\eqref{eq:data2-c}.

    \item \textbf{($(N,\kappa)$ satisfies condition~\eqref{eq:data3-c})} Since $L_3^{\mathrm s}=0$, every summand in~\eqref{eq:loss3_c} is zero. Therefore, whenever
\[
\max_{k\in[K-1]} N(f^k(\underline{x}^i,\kappa(\underline{x}^i)))\le \delta,
\]
we have
\[
N(f^K(\bar{x}^i,\kappa(\bar{x}^i)))\le \delta,
\]
which gives~\eqref{eq:data3-c}.
\end{enumerate}
This implies that, $(N,\kappa)$ satisfies conditions~\eqref{eq:data1-c}--\eqref{eq:data3-c}. Now we can use Theorem~\ref{thm:datamonbar}~part~\ref{thm:datamonbar1} to deduce that the system $\mathfrak S$ is safely controllable with respect to $\Xx_u$.

Regarding part~\ref{thm:zeroloss2}, our goal is to show that if there exists a monotone neural network
$N:\Xx\to\real$ such that
\[
L_{\mathrm{total}}^{\mathrm v}
=
L_1^{\mathrm v}+L_2^{\mathrm v}+L_3^{\mathrm v}=0,
\]
then the system $\mathfrak S$ is robustly safe with respect to $\Xx_u$.
Since each loss term is nonnegative, we obtain
\[
L_1^{\mathrm v}=L_2^{\mathrm v}=L_3^{\mathrm v}=0.
\]
We first show that $N$ satisfies conditions~\eqref{eq:data1}--\eqref{eq:data3}.

\begin{enumerate}

\item \textbf{($N$ satisfies condition~\eqref{eq:data1})}
Since $L_1^{\mathrm v}=0$, every summand
in~\eqref{eq:loss1} is zero. Hence, for every $i\in\mathcal I$ and every
$\overline{\mathbf u}\in\overline{\mathcal U}$,
\[
\max_{k\in[K-1]}
N(f^k(\bar{x}^i,\overline{\mathbf u}))
\le \gamma,
\]
which gives~\eqref{eq:data1}.

\item \textbf{($N$ satisfies condition~\eqref{eq:data2})} Since $L_2^{\mathrm v}=0$, every summand in~\eqref{eq:loss2} must be zero. Hence, for every $i\in\mathcal U$,
\[
\Relu(\eta-N(\underline{x}^i))=0.
\]
Thus,
$N(\underline{x}^i)\ge \eta$.
Hence, condition~\eqref{eq:data2} holds.

\item \textbf{($N$ satisfies condition~\eqref{eq:data3})}
Since $L_3^{\mathrm v}=0$, every summand
in~\eqref{eq:loss3} is zero. Let
$i\in\mathcal P$ and
$\underline{\mathbf u}\in\underline{\mathcal U}$
be such that
\[
\max_{k\in[K-1]}
N(f^k(\underline{x}^i,\underline{\mathbf u}))
\le\delta.
\]
By the definition of $r_i$, we have $r_i=1$.
Thus, for every
$\overline{\mathbf u}\in\overline{\mathcal U}$,
\[
\Relu\!\left(
N(f^K(\bar{x}^i,\overline{\mathbf u}))
-\delta
\right)=0,
\]
which implies
\[
N(f^K(\bar{x}^i,\overline{\mathbf u}))
\le\delta.
\]
Hence, condition~\eqref{eq:data3} holds.

\end{enumerate}
This implies that, $N$ satisfies conditions~\eqref{eq:data1}--\eqref{eq:data3}. Now we can use Theorem~\ref{thm:datamonbar}~part~\ref{thm:datamonbar2} to deduce that the system $\mathfrak S$ is robustly safe with respect to $\Xx_u$. This completes the proof.

\clearpage
\section{Algorithms}
\label{App:algorithms}

\subsection{Monotone Neural Robust Barrier Certificate}
\label{App:alg:verification}

\begin{algorithm}[!h]
\caption{Monotone neural robust barrier learning with adaptive refinement}
\label{alg:learningrobustwithshrinking}

\textbf{Inputs:} Simulator access to the dynamics $f$; hyper-rectangular covers
$\{[\underline{x}^{i},\bar{x}^{i}]\}_{i\in\mathcal P}$, $\{[\underline{x}_0^i,\bar{x}_0^i]\}_{i\in\mathcal I}$, and
$\{[\underline{x}_u^j,\bar{x}_u^j]\}_{j\in\mathcal U}$,
of $\Xx$, $\Xx_0$, and $\Xx_u$, respectively;
an input cover $\{[\underline{u}^{\,q},\bar{u}^{\,q}]\}_{q\in\mathcal Q}$ of $U$;
margins $\gamma<\delta<\eta$; maximum state-refinement budget $n_{\max}^{\mathcal P}$;
maximum input-refinement budget $n_{\max}^{\mathcal Q}$; maximum inductive parameter $K_{\max}$; maximum number of epochs $N_{\mathrm{ep}}$; fixed per-coordinate search budget $N_{\mathrm{search}}\in\mathbb{N}_{>0}$;
and constants $\epsilon>0$ and $C_{\max}>1+\epsilon$.

\textbf{Output:} A monotone neural robust barrier certificate $N$ and final loss
$L_{\mathrm{total}}^{\mathrm v}$, if found; otherwise \texttt{FAIL} and final loss
$L_{\mathrm{total}}^{\mathrm v}$.

\begin{algorithmic}[1]

\State Initialize
$\mathcal H \gets \{[\underline{x}^{\,\ell},\bar{x}^{\,\ell}]\}_{\ell\in\mathcal P \setminus \mathcal{U}},
\;\;
\mathcal J \gets \{[\underline{u}^{\,q},\bar{u}^{\,q}]\}_{q\in\mathcal Q},
\;\;
\mathcal C\gets\emptyset$,\;\;
$L_{\mathrm{total}}^{\mathrm v}\gets\infty$.

\For{$p=1,\ldots,n_{\max}^{\mathcal P}$}

    \For{$K=1,\ldots,K_{\max}$}

        \State Define
        $L_{\mathrm{shrink}}
        =
        \sum_{x^c\in\mathcal C}
        \big[\Relu(\eta-N(x^c))\big]^2$.

        \State Train $N$ for at most $N_{\mathrm{ep}}$ epochs using
        $L_{\mathrm{total}}^{\mathrm v}
        =
        L_1^{\mathrm v}+L_2^{\mathrm v}+L_3^{\mathrm v}
        +L_{\mathrm{shrink}}$,
        where
        \Statex \hspace{\algorithmicindent}\hspace{\algorithmicindent}
        $L_1^{\mathrm v},L_2^{\mathrm v},L_3^{\mathrm v}$ are defined in~\eqref{eq:loss1}--\eqref{eq:loss3}
        using $\mathcal I$, $\mathcal U$, the current input-cover index set induced 
        \Statex \hspace{\algorithmicindent}\hspace{\algorithmicindent}
        by $\mathcal J$, the current unresolved index set induced by $\mathcal H$, the current value of $K$, and margins
        \Statex \hspace{\algorithmicindent}\hspace{\algorithmicindent}$\gamma,\delta,\eta$.

        \If{$L_{\mathrm{total}}^{\mathrm v}=0$}
            \State \textbf{return} $N,L_{\mathrm{total}}^{\mathrm v}$
        \EndIf

    \EndFor

    \State Apply either Algorithm~\ref{alg:monotoneshrinking}
    with budget $N_{\mathrm{search}}$ or
    Algorithm~\ref{alg:arbgrid} with parameters $\epsilon,C_{\max}$ to obtain $(\mathcal H^{+},\mathcal C^{+})$

    \If{the chosen state-refinement algorithm returns \texttt{FAIL}}
        \State \textbf{return} \texttt{FAIL}, $L_{\mathrm{total}}^{\mathrm v}$
    \Else
        \State Update $(\mathcal H,\mathcal C)$ using the returned $(\mathcal H^{+},\mathcal C^{+})$
    \EndIf

    \If{$2|\mathcal J|\le n_{\max}^{\mathcal Q}$}
        \State Apply Algorithm~\ref{alg:inputgrid} to obtain a refined input cover $\mathcal J^{+}$
        \State Update $\mathcal J\gets\mathcal J^{+}$
    \EndIf

\EndFor

\State \textbf{return} \texttt{FAIL}, $L_{\mathrm{total}}^{\mathrm v}$

\end{algorithmic}
\end{algorithm}


\subsection{Monotone Neural Control Barrier Certificate}
\label{App:alg:control}
Algorithm~\ref{alg:learningwithshrinking} presents the control barrier learning procedure. To ensure that the learned controller satisfies the input constraints $U=[u_{\min},u_{\max}]$, we parameterize it as $\kappa(x)=u_c+u_r\odot\tanh(\tilde{\kappa}(x))$, where $u_c=(u_{\max}+u_{\min})/2$,
$u_r=(u_{\max}-u_{\min})/2$, and
$\tilde{\kappa}$ is the Max-Monotone Network controller. Since $\tanh$ is nondecreasing and $u_r\ge 0$,
the resulting controller remains monotone and satisfies
$\kappa(x)\in U$ for all $x\in\Xx$.

\clearpage
\subsection{Monotone-Shrinking Hyper-Rectangular Search}
\label{App:gridtrain}

\begin{algorithm}[!h]
\caption{Monotone shrinking of hyper-rectangular grid}
\label{alg:monotoneshrinking}

\textbf{Inputs:} A trained monotone neural certificate $N$; the current unresolved hyper-rectangular collection $\mathcal H=\{[\underline{x}^{\,\ell},\bar{x}^{\,\ell}]\}_{\ell\in\mathcal P \setminus \mathcal{U}}$; the current shrinking-constraint set $\mathcal C$; a fixed per-coordinate search budget $N_{\mathrm{search}}\in\mathbb N_{>0}$.

\textbf{Output:} A refined unresolved collection $\mathcal H^{+}$ and an updated shrinking-constraint set $\mathcal C^{+}$; or \texttt{FAIL}.

\begin{algorithmic}[1]

\State Initialize $\mathcal H^{+}\gets \emptyset$ and $\mathcal C^{+}\gets \mathcal C$

\For{each box $[\underline{x}^{\,\ell},\bar{x}^{\,\ell}]\in\mathcal H$}
    \For{$i=1,\ldots,n$}
        \State Using at most $N_{\mathrm{search}}$ evaluations of $N$, attempt to find $\xi_i^\ell\in[\underline{x}^{\,\ell}_i,\bar{x}^{\,\ell}_i]$ such that
        \[
        x^{c,\ell,i}
        =
        (\underline{x}^{\,\ell}_1,\ldots,\underline{x}^{\,\ell}_{i-1},
        \xi_i^\ell,
        \underline{x}^{\,\ell}_{i+1},\ldots,\underline{x}^{\,\ell}_n)
        \quad\text{satisfies}\quad
        N(x^{c,\ell,i}) > 0 .
        \]
        \If{such a point is found}
            \State $\mathcal C^{+}\gets \mathcal C^{+}\cup\{x^{c,\ell,i}\}$
        \Else
            \State $\xi_i^\ell\gets \bar{x}^{\,\ell}_i$
        \EndIf
    \EndFor

    \State Define $\hat{x}^{\,\ell}=(\xi_1^\ell,\ldots,\xi_n^\ell)$

    \If{$\hat{x}^{\,\ell}=\bar{x}^{\,\ell}$}
        \State \textbf{return} \texttt{FAIL}
    \EndIf

    \State $\mathcal H^{+}\gets \mathcal H^{+}\cup\{[\underline{x}^{\,\ell},\hat{x}^{\,\ell}]\}$

\EndFor

\State \textbf{return} $\mathcal H^{+},\mathcal C^{+}$

\end{algorithmic}
\end{algorithm}

Since each coordinate-wise search has a fixed budget, Algorithm~\ref{alg:monotoneshrinking} performs at most $N_{\mathrm{search}}n|\mathcal H|$ neural-network evaluations per refinement round.

\clearpage
\subsection{Arbitrary Hyper-Rectangular Gridding}\label{App:grid}

\begin{algorithm}[!h]
\caption{Arbitrary refinement of state space grid}
\label{alg:arbgrid}

\textbf{Inputs:} The current unresolved hyper-rectangular collection
$\mathcal H=\{[\underline{x}^{\,\ell},\bar{x}^{\,\ell}]\}_{\ell\in\mathcal P \setminus \mathcal{U}}$;
the current shrinking-constraint set $\mathcal C$; and constants
$\epsilon>0$ and $C_{\max}>1+\epsilon$ defining the random split range.

\textbf{Output:} A refined unresolved collection $\mathcal H^{+}$ and an unchanged shrinking-constraint set $\mathcal C^{+}$.

\begin{algorithmic}[1]

\State Initialize $\mathcal H^{+}\gets \emptyset$ and $\mathcal C^{+}\gets \mathcal C$

\For{each box $[\underline{x}^{\,\ell},\bar{x}^{\,\ell}]\in\mathcal H$}

    \State Let $d=\arg\max_k(\bar{x}^{\,\ell}_k-\underline{x}^{\,\ell}_k)$
    \State Sample $r_x\sim\mathrm{Uniform}(1+\epsilon,C_{\max})$
    \State Compute
    $c_d^\ell
    =
    \underline{x}^{\,\ell}_d
    +
    \frac{\bar{x}^{\,\ell}_d-\underline{x}^{\,\ell}_d}{r_x}$.

    \State Let $\bar{x}^{\,\ell,1}=\bar{x}^{\,\ell}$ with $\bar{x}^{\,\ell,1}_d=c_d^\ell$
    \State Let $\underline{x}^{\,\ell,2}=\underline{x}^{\,\ell}$ with $\underline{x}^{\,\ell,2}_d=c_d^\ell$

    \State Define
    $H_1^\ell=[\underline{x}^{\,\ell},\bar{x}^{\,\ell,1}],
    \;\;
    H_2^\ell=[\underline{x}^{\,\ell,2},\bar{x}^{\,\ell}]$.

    \State $\mathcal H^{+}\gets \mathcal H^{+}\cup\{H_1^\ell,H_2^\ell\}$

\EndFor

\State \textbf{return} $\mathcal H^{+},\mathcal C^{+}$

\end{algorithmic}
\end{algorithm}

\begin{algorithm}[!h]
\caption{Arbitrary refinement of input space grid}
\label{alg:inputgrid}

\textbf{Inputs:} The current input cover
$\mathcal J$; and constants $\epsilon>0$ and $C_{\max}>1+\epsilon$ defining the random split range.

\textbf{Output:} A refined input cover $\mathcal J^{+}$.

\begin{algorithmic}[1]

\State Initialize $\mathcal J^{+}\gets \emptyset$

\For{each box $[\underline{u}^{\,q},\bar{u}^{\,q}]\in\mathcal J$}

    \State Let $d=\arg\max_k(\bar{u}^{\,q}_k-\underline{u}^{\,q}_k)$
    \State Sample $r_u\sim\mathrm{Uniform}(1+\epsilon,C_{\max})$
    \State Compute
    $c_d^q
    =
    \underline{u}^{\,q}_d
    +
    \frac{\bar{u}^{\,q}_d-\underline{u}^{\,q}_d}{r_u}$.

    \State Let $\bar{u}^{\,q,1}=\bar{u}^{\,q}$ with $\bar{u}^{\,q,1}_d=c_d^q$
    \State Let $\underline{u}^{\,q,2}=\underline{u}^{\,q}$ with $\underline{u}^{\,q,2}_d=c_d^q$

    \State Define
    $J_1^q=[\underline{u}^{\,q},\bar{u}^{\,q,1}],
    \;\;
    J_2^q=[\underline{u}^{\,q,2},\bar{u}^{\,q}]$.

    \State $\mathcal J^{+}\gets \mathcal J^{+}\cup\{J_1^q,J_2^q\}$

\EndFor

\State \textbf{return} $\mathcal J^{+}$

\end{algorithmic}
\end{algorithm}

\clearpage
\section{The role of the inductive parameter $K$ in sample complexity}\label{App:k}
\begin{figure}[!h]
 \centering
  \includegraphics[width =0.6\linewidth]{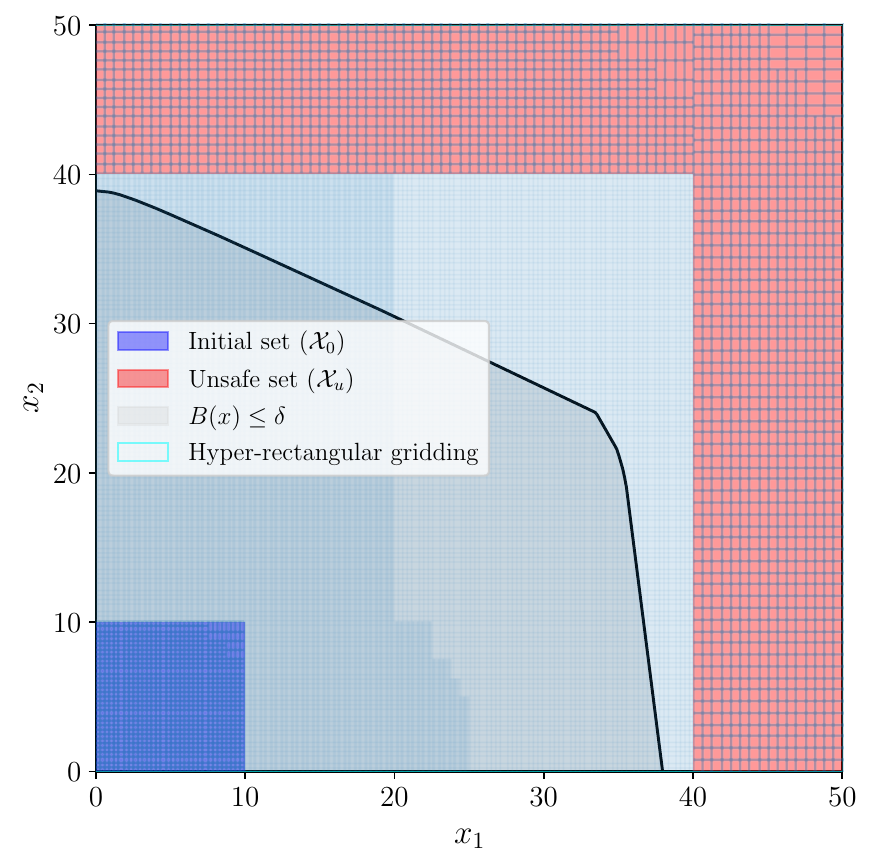}
  \caption{Sub-level set of the learned neural barrier certificate (one-step barrier) over the state space. The parameters $\gamma = 5$, $\delta = 9.5$, and $\eta = 10$ are used in this experiment.}
	\label{fig:fullgrid}
 \end{figure}

In this section, we investigate the effect of the inductive parameter $k$ on the sample complexity. To this end, we consider the 2-D traffic system described in Appendix~\ref{App:compbarr}. As shown previously, our method verifies safety for this system using an extremely coarse rectangular gridding (two rectangles covering the unsafe region, one rectangle for the initial set, and one rectangle for the remaining safe region), enabled by the long inductive parameter $k=100$. In contrast, for the same coarse gridding, ordinary one-step barrier conditions (based only on extreme points and a single-step horizon) are insufficient to certify safety. Motivated by this limitation, we now refine the gridding over the state space using Algorithm~\ref{alg:arbgrid} and try to learn a classic neural barrier certificate from the resulting denser set of samples. Figure~\ref{fig:fullgrid} shows that, for a classic (one-step) barrier certificate (i.e., the inductive parameter $K=1$), safety verification requires a very dense rectangular gridding ($103{,}000$ boxes), which leads to a sample-intensive procedure and becomes impractical for high-dimensional systems. In this experiment, we use a simplified Max-Monotone Network based on the direct max-input branch in Definition~\ref{eq:monnn}, with two hidden layers of widths $200$ and $100$, respectively, and Leaky ReLU activations $(\alpha=0.1)$.

As illustrated in Figure~\ref{fig:fullgrid}, without the benefit of an inductive barrier (even though we still exploit monotonicity of both the system and the barrier function), the required sample complexity remains high, even for this simple 2-D case study. This highlights the crucial role of the inductive parameter $K$ in mitigating sample complexity.

\section{Proof of Theorem~\ref{thm:shrinkcontrolbarrier}}
\label{App:proof_shrinkcontrol}

The proof of Theorem~\ref{thm:shrinkcontrolbarrier} is provided below.

\begin{proof}
We divide the proof into two parts.

\medskip
\noindent\textit{(i) Correctness.}
Suppose Algorithm~\ref{alg:learningwithshrinking}, using Algorithm~\ref{alg:monotoneshrinking} as the refinement routine, returns $(N,\kappa,L_{\mathrm{total}})$. By the termination condition of the algorithm,
$L_{\mathrm{total}} = L_1 + L_2 + L_3 + L_{\mathrm{shrink}} = 0$.
Since all loss terms are nonnegative, it follows that
\[
L_1=L_2=L_3=L_{\mathrm{shrink}}=0.
\]
By the standing result already established in the paper, $L_1=L_2=L_3=0$ implies that $(N,\kappa)$ satisfies the $K$-inductive control barrier conditions~\eqref{eq:data1-c}--\eqref{eq:data3-c}. Moreover,
$L_{\mathrm{shrink}}=0$
implies
\[
N(x^{c,\ell,i}_{(q)}) \ge \eta > 0
\qquad
\forall\, x^{c,\ell,i}_{(q)}\in\mathcal C,
\]
so all shrinking constraints generated in previous iterations are preserved and no previously certified shrinking step is invalidated by retraining. Therefore, the returned pair $(N,\kappa)$ satisfies the required control barrier conditions together with all accumulated shrinking constraints. The safety of $\Sigma$ with respect to $\Xx_u$ then follows from the control barrier theorem.

\medskip
\noindent\textit{(ii) Linear simulator-query complexity.}
Let $M:=|\mathcal P_0|$. When Algorithm~3 is used, each unresolved box produces one refined box, so $|\mathcal H|\le M$, independently of $n$. Each refinement round requires at most $N_{\mathrm{search}}Mn$ evaluations of the
learned certificate $N$, but these evaluations do not require simulator queries.

Simulator queries are instead used to evaluate the trajectories appearing in $L_1,L_2,L_3$. Since the numbers of initial, unsafe, and unresolved cells are bounded independently of $n$, each loss evaluation involves a bounded number of trajectories, each requiring at most
$K_{\max}$ simulator queries. Moreover, the
numbers of refinement rounds, inductive-parameter updates, and training epochs are bounded by $n_{\max}^{\mathcal P}$, $K_{\max}$, and $N_{\mathrm{ep}}$, respectively.

Assuming a fixed number of simulator evaluations per trajectory during training, these bounds are independent of $n$. Thus, the total number of
simulator queries is bounded independently of $n$, and in particular is $O(n)$.
\end{proof}

\section{Correctness and Linear Simulator-Query Complexity for Verification}
\label{App:ver_On}

\begin{theorem}[Correctness and linear simulator-query complexity of Algorithm~\ref{alg:learningrobustwithshrinking}]
\label{thm:shrinkrobustbarrier}
Consider a monotone discrete-time system
$\Sigma=(\Xx,\Xx_0,U,f), \; \Xx\subseteq\mathbb R^n$,
and suppose that
\[
\Xx_0 \subseteq \bigcup_{i\in\mathcal I}[\underline{x}_0^i,\bar{x}_0^i],
\qquad
\Xx_u \subseteq \bigcup_{j\in\mathcal U}[\underline{x}_u^j,\bar{x}_u^j],
\qquad
\Xx\setminus\Xx_u
\subseteq
\bigcup_{\ell\in\mathcal P_0}[\underline{x}^{\,\ell},\bar{x}^{\,\ell}],
\]
and
\[
U\subseteq
\bigcup_{q\in\mathcal Q_0}[\underline{u}^{\,q},\bar{u}^{\,q}],
\]
for finite index sets $\mathcal I,\mathcal U,\mathcal P_0$, and $\mathcal Q_0$ whose cardinalities are independent of $n$. Then:
\begin{enumerate}
    \item[\textnormal{(i)}] If Algorithm~\ref{alg:learningrobustwithshrinking}, using either Algorithm~\ref{alg:monotoneshrinking} or Algorithm~\ref{alg:arbgrid} as the state-refinement routine and Algorithm~\ref{alg:inputgrid} as the input-refinement routine, returns $(N,L_{\mathrm{total}}^{\mathrm v})$, then the system $\Sigma$ is robustly safe with respect to $\Xx_u$.
    
    \item[\textnormal{(ii)}] The total number of simulator queries performed by Algorithm~\ref{alg:learningrobustwithshrinking}, using Algorithm~\ref{alg:monotoneshrinking} as the state-refinement routine and Algorithm~\ref{alg:inputgrid} as the input-refinement routine, is $\mathcal O(n)$.
\end{enumerate}
\end{theorem}

\begin{proof}
We divide the proof into two parts.

\medskip
\noindent\textit{(i) Correctness.}
Suppose Algorithm~\ref{alg:learningrobustwithshrinking}, using either Algorithm~\ref{alg:monotoneshrinking} or Algorithm~\ref{alg:arbgrid} as the state-refinement routine and Algorithm~\ref{alg:inputgrid} as the input-refinement routine, returns $(N,L_{\mathrm{total}}^{\mathrm v})$. By the termination condition of the algorithm,
\[
L_{\mathrm{total}}^{\mathrm v}
=
L_1^{\mathrm v}+L_2^{\mathrm v}+L_3^{\mathrm v}
+L_{\mathrm{shrink}}
=0.
\]
Since all loss terms are nonnegative, it follows that
\[
L_1^{\mathrm v}
=
L_2^{\mathrm v}
=
L_3^{\mathrm v}
=
L_{\mathrm{shrink}}
=
0.
\]

By Theorem~\ref{thm:zeroloss},
$L_1^{\mathrm v}=L_2^{\mathrm v}=L_3^{\mathrm v}=0$
implies that $N$ satisfies the $K$-inductive robust barrier conditions~\eqref{eq:data1}--\eqref{eq:data3}. Moreover,
$L_{\mathrm{shrink}}=0$
implies
\[
N(x^{c,\ell,i}_{(q)}) \ge \eta > 0,
\qquad
\forall\, x^{c,\ell,i}_{(q)}\in\mathcal C,
\]
so all accumulated shrinking constraints, if any, are preserved throughout the training process. Therefore, the returned certificate $N$ satisfies the required robust barrier conditions together with all accumulated shrinking constraints. The robust safety of $\Sigma$ with respect to $\Xx_u$ then follows from the robust barrier theorem.

\medskip
\noindent\textit{(ii) Linear simulator-query complexity under monotone shrinking refinement.}
Let $M:=|\mathcal P_0|$. When Algorithm~\ref{alg:monotoneshrinking} is used as the state-refinement routine, each unresolved box produces one refined unresolved box. Hence, the number of unresolved boxes remains bounded by $M$, independently of $n$.

For each unresolved box and each coordinate $i\in\{1,\ldots,n\}$, Algorithm~\ref{alg:monotoneshrinking} performs a one-dimensional search using at most $N_{\mathrm{search}}$ evaluations of the learned certificate $N$. Thus, each refinement round requires at most $N_{\mathrm{search}}Mn$ neural-network evaluations, which do not involve simulator queries.

Simulator queries are instead required to evaluate the trajectories appearing in the loss terms $L_1^{\mathrm v}$, $L_2^{\mathrm v}$, and $L_3^{\mathrm v}$. Since the cardinalities of $\mathcal I$, $\mathcal U$, and $\mathcal P_0$ are independent of $n$, and the input-refinement budget $n_{\max}^{\mathcal Q}$ is fixed, the number of trajectories evaluated per loss computation is bounded independently of $n$. Each trajectory requires at most $K_{\max}$ simulator evaluations.

Moreover, the numbers of refinement rounds, inductive-parameter updates, and training epochs are bounded by $n_{\max}^{\mathcal P}$, $K_{\max}$, and $N_{\mathrm{ep}}$, respectively. Assuming that each training epoch uses a bounded number of trajectory evaluations without additional dimension-dependent simulator queries, the total number of simulator queries is bounded independently of $n$, and hence is $\mathcal O(n)$.
\end{proof}

\section{Numerical Experiment Details}
\label{App:models}

In this appendix, we provide details of our numerical experiments in Section~\ref{sec:numericalexp}. 

\subsection{Baseline methods and comparisons}\label{app:baseline}

The baseline methods considered for comparison are Method~1, the counterexample-guided approach~\cite{edwards2024fossil}, and Method~2, the Lipschitz-based method~\cite{anand2023formally}. Although~\cite{edwards2024fossil} requires explicit access to the dynamics of the system during its formal guarantee stage, we include it for comparison since its barrier construction is still data-driven. In particular,~\cite{edwards2024fossil} relies on fixed random sampling and SMT-based verification, which fails to provide guarantees as the dimension grows. 

All experiments of the proposed method were conducted on a MacBook Pro equipped with an Apple M3 Max processor and 48\,GB RAM. All experiments for the method of~\cite{edwards2024fossil} were conducted on a workstation running Ubuntu 20.04, equipped with Intel Xeon Gold 6238R CPUs, two NVIDIA Quadro RTX 8000 GPUs, and 160\,GB of system memory.

In all experiments, the barrier certificate $N:\mathbb{R}^n\to\mathbb{R}$, and in the safe synthesis setting also the feedback controller $\kappa:\mathbb{R}^n\to\mathbb{R}^n$, are parameterized by the Max-Monotone Network architecture with three branches: one direct max-input branch and two affine max branches with projection dimensions $m_1=0.2\,n$ and $m_2=0.1\,n$, except for the Lotka--Volterra experiments reported in Table~\ref{tb:trf-sc}, where we use the fixed dimensions $m_1=10$ and $m_2=20$. Each branch is followed by a monotone subnetwork with two hidden layers of $200$ neurons and Leaky ReLU activations $(\alpha=0.1)$.

\subsection{Freeway Traffic Flow Model}
\label{App:traffic}

\begin{figure}[!b]
\centering
\includegraphics[width=0.7\linewidth]{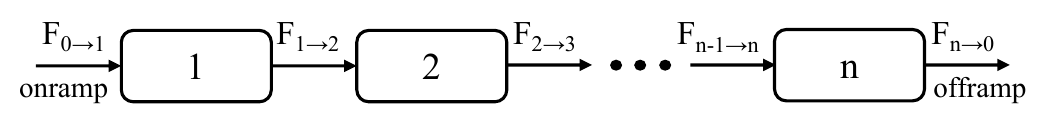}
\caption{A line digraph illustrating an $n$-dimensional traffic network.}
\label{fig:celltran}
\end{figure}

\paragraph{Model.}
We consider the freeway traffic network $\Sys=(\Xx,\Xx_0,U,f)$ illustrated in Fig.~\ref{fig:celltran}. The state is $q(t)=[q_1(t),\ldots,q_n(t)]^\top \in \R^n$, where $q_i(t)\ge 0$ denotes the vehicle density in segment $i$. The state space is $\Xx=[0,50]^n$.

The dynamics $f:\Xx\times U \to \Xx$ are given by
\begin{align}
q_1(t+1) &= q_1(t) + \tau\big(F_{0\to 1}-F_{1\to 2}(q(t))\big),\\
q_i(t+1) &= q_i(t) + \tau\big(F_{i-1\to i}(q(t))-F_{i\to i+1}(q(t))\big),\;\; i=2,\ldots,n-1,\\
q_n(t+1) &= q_n(t) + \tau\big(F_{n-1\to n}(q(t))-F_{n\to 0}(q(t))\big).
\end{align}
The input corresponds to the on-ramp inflow $F_{0\to 1}\in U=[0,7.5]$, with sampling time $\tau=0.5$.

The inter-segment flow is defined by
\begin{align}
F_{i\to i+1}(q) &= \min\{\varphi(q_i),\,\sigma(q_{i+1})\},\\
F_{n\to 0}(q) &= \varphi(q_n),
\end{align}
where
\[
\varphi(x)=15\big(1-e^{-0.07x}\big), \qquad
\sigma(x)=\max\!\left\{0,\;15\Big(1-\big(\tfrac{x}{50}\big)^3\Big)\right\}.
\]

The initial set is $\Xx_0=[0,5]^n$ and the unsafe set is $\Xx_u=\Xx\setminus[0,40]^n$.

\paragraph{Monotonicity.}
The system is monotone with respect to the standard partial order; see~\citep{FB-CTDS} for a detailed proof.

\subsection{Kuramoto Oscillator Networks}
\label{App:kuramoto}

\paragraph{Fully connected network.}
We consider a discrete-time Kuramoto oscillator network
$\Sys=(\Xx,\Xx_0,U,f)$ consisting of $n$ oscillators with fully connected coupling~\cite{skardal2015control}. Let
$\theta(t)=[\theta_1(t);\ldots;\theta_n(t)]\in\R^n$ denote the vector of oscillator phases at time $t$, where $\theta_i(t)\in[0,\tfrac{\pi}{4}]$.

The state, initial, and input sets are
\[
\Xx=\left[0,\tfrac{\pi}{4}\right]^n,\quad
\Xx_0=\left[0,\tfrac{\pi}{16}\right]^n,\quad
U=\left[-\tfrac{\pi}{4},\tfrac{\pi}{4}\right]^n.
\]

The dynamics are given by
\[
\theta_i(t+1)
=
\theta_i(t)
+
\frac{\tau K_c}{n}
\sum_{j\neq i}\sin\!\big(\theta_j(t)-\theta_i(t)\big)
+
\tau u_i(t),
\]
where $\tau>0$ is the sampling time and $K_c>0$ is the coupling strength. In our implementation, the state update is followed by componentwise clipping to $\Xx=[0,\pi/4]^n$, ensuring state-space invariance while preserving monotonicity.

\paragraph{Monotonicity.}
For $j\neq i$, we have
\[
\frac{\partial f_i}{\partial \theta_j}
=
\frac{\tau K_c}{n}\cos(\theta_j-\theta_i)\ge 0,
\]
since $\theta_j-\theta_i\in[-\tfrac{\pi}{4},\tfrac{\pi}{4}]$.

For the diagonal term,
\[
\frac{\partial f_i}{\partial \theta_i}
=
1-\frac{\tau K_c}{n}\sum_{j\neq i}\cos(\theta_j-\theta_i)
\ge 1-\tau K_c.
\]

Thus, if $\tau K_c\le 1$, all entries of the Jacobian are nonnegative, and the system is monotone. Moreover, $\frac{\partial f_i}{\partial u_i}=\tau>0$, so the system is also monotone with respect to the input.

\medskip

\paragraph{Networks with structured coupling (IEEE testbeds).}
We also consider Kuramoto networks with non-uniform coupling structures derived from IEEE power-system testbeds. In this case, the dynamics are given by
\[
\theta_i(t+1)
=
\theta_i(t)
+
\tau u_i(t)
+
\frac{\tau K_c}{|\mathcal{N}_i|}
\sum_{j\in\mathcal{N}_i}
a_{ij}\sin\!\big(\theta_j(t)-\theta_i(t)\big),
\]
where $\mathcal{N}_i$ denotes the set of neighbors of node $i$ and $a_{ij}\in\{0,1\}$ indicates connectivity.

The state, initial, input, and unsafe sets are identical to the fully connected case.

\paragraph{Monotonicity.}
For $j\neq i$,
\[
\frac{\partial f_i}{\partial \theta_j}
=
\frac{\tau K_c}{|\mathcal{N}_i|}
a_{ij}\cos(\theta_j-\theta_i)\ge 0.
\]

For the diagonal term,
\[
\frac{\partial f_i}{\partial \theta_i}
=
1-\frac{\tau K_c}{|\mathcal{N}_i|}
\sum_{j\in\mathcal{N}_i} a_{ij}\cos(\theta_j-\theta_i)
\ge 1-\tau K_c.
\]

Thus, under the same condition $\tau K_c\le 1$, the system is monotone. As before, $\frac{\partial f_i}{\partial u_i}=\tau>0$, so monotonicity with respect to the input also holds.

\subsection{Lotka--Volterra Population Dynamics}
\label{App:lvmon}

\paragraph{Model.}
We consider the $n$-dimensional Lotka--Volterra population model in discrete time, modeled as the system $\mathfrak{S}=(\Xx,\Xx_0,f)$ with state $x(t)\in\R_{\ge 0}^{n}$. Let $A=[a_{ij}]\in\R^{n\times n}$ denote the interaction matrix and let $r\in\R^n$ be the intrinsic growth-rate vector. The discrete-time dynamics are given by
\[
x(t+1)=f(x(t)):=x(t)+\tau\,[x(t)]\big(Ax(t)+r\big),
\]
where $\tau>0$ is the sampling time and $[x(t)]:=\mathrm{diag}(x_1(t),\ldots,x_n(t))$. Equivalently, for each $i\in\{1,\ldots,n\}$,
\[
x_i(t+1)=x_i(t)+\tau\,x_i(t)\Big(\sum_{j=1}^{n} a_{ij}x_j(t)+r_i\Big).
\]

\paragraph{Monotonicity.}
A sufficient condition for $f$ to be monotone (with respect to the standard partial order) on a set $\Xx$ is
\[
Df(x)\ge 0, \qquad \forall x\in\Xx,
\]
where the inequality is understood elementwise.

The Jacobian of $f$ is given by
\[
Df(x)=I+\tau\Big([x]A+\mathrm{diag}(Ax+r)\Big).
\]

For $i\neq j$, the off-diagonal entries are
\[
\frac{\partial f_i}{\partial x_j}(x)=\tau x_i a_{ij}\ge 0,
\]
which holds for all $x\ge 0$ whenever $A$ is Metzler, i.e., $a_{ij}\ge 0$ for all $i\neq j$.

For the diagonal terms,
\[
\frac{\partial f_i}{\partial x_i}(x)
=
1+\tau\Big(x_i a_{ii}+(Ax+r)_i\Big),
\]
and thus we require
\[
1+\tau\Big(x_i a_{ii}+(Ax+r)_i\Big)\ge 0,
\qquad \forall x\in\Xx.
\]

Now consider the hyper-rectangle $\Xx=[0,\bar{x}]^n$. Since $A$ is Metzler and $x\ge 0$, we have
\[
(Ax+r)_i
= r_i + a_{ii}x_i + \sum_{j\neq i} a_{ij}x_j
\ge
r_i + a_{ii}x_i,
\]
which implies
\[
x_i a_{ii}+(Ax+r)_i
\ge
r_i + 2a_{ii}x_i.
\]

Assuming $a_{ii}\le 0$ for all $i$, we further obtain
\[
r_i + 2a_{ii}x_i \ge r_i + 2a_{ii}\bar{x}.
\]

Therefore, a sufficient condition for monotonicity on $\Xx=[0,\bar{x}]^n$ is
\[
1+\tau\big(r_i+2a_{ii}\bar{x}\big)\ge 0,
\qquad \forall i\in\{1,\ldots,n\}.
\]

Equivalently, it suffices to choose $\tau$ such that
\[
\tau \le \min_{i:\,r_i+2a_{ii}\bar{x}<0}
\frac{1}{-\big(r_i+2a_{ii}\bar{x}\big)},
\]
with the convention that the bound is $+\infty$ if
$r_i+2a_{ii}\bar{x}\ge 0$ for all $i$.

\subsection{Ribosome Flow Model}
\label{App:rfm}
\paragraph{Model.}
We consider a Ribosome Flow Model (RFM)
$\mathfrak{S}=(\Xx,\Xx_0,f)$ consisting of $n$ sites. Let
$x(t)=[x_1(t);\ldots;x_n(t)]$ denote the state at time step $t$, where
$x_i(t)\in[0,1]$ represents the occupancy level of site $i$. The dynamics are
given by
\begin{align*}
x_1(t+1) &= x_1(t) + h\!\left(\lambda_0(1-x_1(t)) - \lambda_1 x_1(t)(1-x_2(t))\right), \\
x_i(t+1) &= x_i(t) + h\!\left(\lambda_{i-1}x_{i-1}(t)(1-x_i(t)) - \lambda_i x_i(t)(1-x_{i+1}(t))\right), 
\quad i=2,\ldots,n-1, \\
x_n(t+1) &= x_n(t) + h\!\left(\lambda_{n-1}x_{n-1}(t)(1-x_n(t)) - \lambda_n x_n(t)\right).
\end{align*}
where $\lambda_i>0$ are transition rates and $h>0$ is the sampling time.

In our experiments, we sample $\lambda_i$ independently
from the uniform distribution $\lambda_i\sim\mathcal{U}(0.8,1.2)$. The step size
$h$ is chosen to satisfy the monotonicity condition
\[
h\leq \big(\max_i(\lambda_i+\lambda_{i+1})\big)^{-1}.
\]

We consider the state space $\Xx = [0,1]^n$ and the initial set $\Xx_0 = [0,0.2]^n$. 
The unsafe set is defined as
\[
\Xx_u = [0,1]^n \setminus [0,0.9]^n.
\]

\paragraph{Monotonicity.}
The monotonicity of the system follows from the Jacobian of the evaluation map. 
For $j\neq i$, we have
\[
\frac{\partial f_i}{\partial x_j}(x)\geq 0,
\]
since all off-diagonal terms are of the form
$h\lambda_i x_i$ or $h\lambda_i(1-x_i)$, which are nonnegative on $[0,1]^n$.

For the diagonal terms,
\[
\frac{\partial f_i}{\partial x_i}(x)
=
1 - h\lambda_{i-1}x_{i-1} - h\lambda_i(1-x_{i+1}),
\]
and since $x_i\in[0,1]$, it follows that
\[
\frac{\partial f_i}{\partial x_i}(x)\geq 1 - h(\lambda_{i-1}+\lambda_i).
\]

Thus, under the condition
\[
h\leq (\lambda_{i-1}+\lambda_i)^{-1}, \quad \forall i,
\]
all entries of the Jacobian are nonnegative, and the system is monotone on $\Xx$.



\end{document}